\documentclass{aa}  
\pdfoutput=1
\usepackage{graphicx}
\usepackage{amsmath}	
\usepackage{amssymb}	
\usepackage{relsize}
\usepackage{txfonts}
\usepackage[caption=false]{subfig}
\usepackage[colorlinks=true,linkcolor=blue,allcolors=blue]{hyperref}%

\newcommand{\Msun}{M$_\odot$}

\begin{document} 

\title{Robust machine learning model of inferring the ex situ stellar fraction of galaxies from photometric data}

\subtitle{}

\author{Runsheng Cai$^{1,2}$\thanks{E-mail: cairunsheng@shao.ac.cn}, Ling Zhu$^2$\thanks{Corr author: lzhu@shao.ac.cn}, Shiyin Shen$^2$, Wenting Wang$^3$, Annalisa Pillepich$^4$, Jes\'us Falc\'on-Barroso$^{5,6}$ }
%, Shiyin Shen, Wenting Wang$^2$, Song Huang$^3$ , Annalisa Pillepich$^4$}

%, Lodovico Coccato$^{10}$, Enrico Maria Corsini$^{4,5}$,  Katja Fahrion$^{10}$,  Jes\'us Falc\'on-Barroso$^{8,9}$, Dimitri A. Gadotti$^{10}$, Enrica Iodice$^{10}$, Mariya Lyubenova$^{10}$, Shude Mao$^{17}$,  Ignacio Martin Navarro$^{8,9}$, Richard McDermid$^{11}$, Francesca Pinna$^{2}$, Adriano Poci$^{11}$, Mark Sarzi$^{12,13}$, Tim de Zeeuw$^{6,7}$

\institute{
School of Astronomy and Space Sciences, University of Chinese Academy of Sciences, No. 19A Yuquan Road, Beijing 100049, People’s Republic of China\\
\and
Shanghai Astronomical Observatory, Chinese Academy of Sciences, 80 Nandan Road, Shanghai 200030, China
%Department of Astrophysics, University of Vienna, T\"urkenschanzstra{\ss}e 17, 1180 Vienna, Austria
\and
Department of Astronomy, Shanghai Jiao Tong University, Shanghai 200240, China
%\and
%Department of Astronomy, Tsinghua University, Beijing 100084, China
\and
Max Planck Institute for Astronomy, K\"onigstuhl 17, 69117 Heidelberg, Germany % MPIA
%\and
%Dipartimento di Fisica e Astronomia 'G. Galilei', Universit\`a di Padova,  vicolo dell'Osservatorio 3, I-35122 Padova, Italy %Enrico
%\and
%INAF--Osservatorio Astronomico di Padova, vicolo dell'Osservatorio 5, I-35122 Padova, Italy % Enrico
%\and
%Sterrewacht Leiden, Leiden University, Postbus 9513, 2300 RA Leiden, The Netherlands % Tim
%\and
%Max-Planck-Institut f\"ur extraterrestrische Physik, Gie\ss{}enbachstra\ss{}e 1, 85748 Garching bei M\"unchen, Germany % Tim
\and
Instituto de Astrof\'isica de Canarias, Calle Via L\'{a}ctea s/n, 38200 La Laguna, Tenerife, Spain% Jesus
\and
Depto. Astrof\'isica, Universidad de La Laguna, Calle Astrof\'isico Francisco S\'{a}nchez s/n, 38206 La Laguna, Tenerife, Spain% Jesus
%\and
%European Southern Observatory, Karl-Schwarzschild-Stra\ss{}e 2, 85748 Garching bei M\"unchen, Germany %ESO
%\and
%Department of Physics and Astronomy, Macquarie University, North Ryde, NSW 2109, Australia %Richard
%\and
%Armagh Observatory and Planetarium, College Hill, Armagh, BT61 9DG, Northern Ireland, UK  % Sarzi
%\and
%Centre for Astrophysics Research, University of Hertfordshire, College Lane, Hatfield AL10 9AB, UK % Sarzi
%\and
%Department of Astronomy and Tsinghua Center for Astrophysics, Tsinghua University, Beijing 100084, China % Dandan, Shude
%\and
%European Space Agency (ESA), European Space Research and Technology Centre (ESTEC), Keplerlaan 1, 2201 AZ Noordwijk,
 % The Netherlands
             }

\date{Received; accepted}
   
\titlerunning{Inferring the ex situ stellar mass fraction}
\authorrunning{Cai et al.}  
 
  \abstract
   {We search for parameters defined from photometric images to quantify the ex situ stellar mass fraction of galaxies.
   We created mock images using galaxies in the cosmological hydrodynamical simulations TNG100, EAGLE, and TNG50 at redshift $z=0$. We define a series of parameters describing their structures, including: the absolute magnitude in $r$ and $g$ bands ($M_r$, $M_g$), the half-light and 90\%-light radius ($r_{50}$, $r_{90}$), the concentration ($C$), the luminosity fractions of inner and outer halos ($f_{\rm innerhalo}$, $f_{\rm outerhalo}$), the inner and outer surface brightness gradients ($\nabla \rho_{\rm inner}$,$\nabla \rho_{\rm outer}$), and \emph{g-r} colour gradients ($\nabla \rm (\emph{g-r})_{inner}$,$\nabla \rm (\emph{g-r})_{outer}$). In particular, the inner and outer halo of a galaxy are defined by sectors ranging from $45-135$ degrees from the disk major axis, and with radii ranging from $3.5-10$ kpc and $10-30$ kpc, respectively, to avoid the contamination of disk and bulge. The surface brightness and colour gradients are defined by the same sectors along the minor axis and with similar radii ranges. We used the Random Forest method to create a model that predicts $f_{\rm exsitu}$ from morphological parameters. The model predicts $f_{\rm exsitu}$ well with a scatter smaller than 0.1 compared to the ground truth in all mass ranges. The models trained from TNG100 and EAGLE work similarly well and are cross-validated; they also
work well in making predictions for TNG50 galaxies. The analysis using Random Forest reveals that $\nabla \rho_{\rm outer}$, $\nabla \rm (\emph{g-r})_{outer}$, $f_{\rm outerhalo}$ and $f_{\rm innerhalo}$ are the
   most influential parameters in predicting $f_{\rm exsitu}$, underscoring their significance in uncovering the merging history of galaxies. We further analyse how the quality of images will affect the results by using SDSS-like and HSC-like mock images for galaxies at different distances. Our results can be used to infer the ex situ stellar mass fractions for a large sample of galaxies from photometric surveys.
   }


   \maketitle
\begin{figure*}
\centering\includegraphics[width=5.6cm]{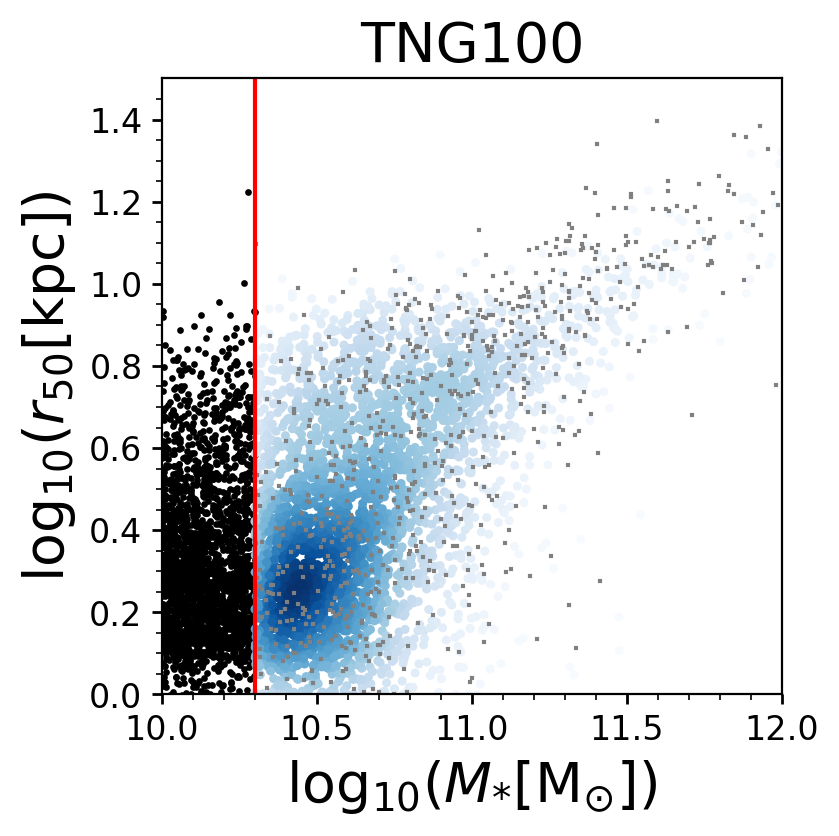}\centering\includegraphics[width=5.6cm]{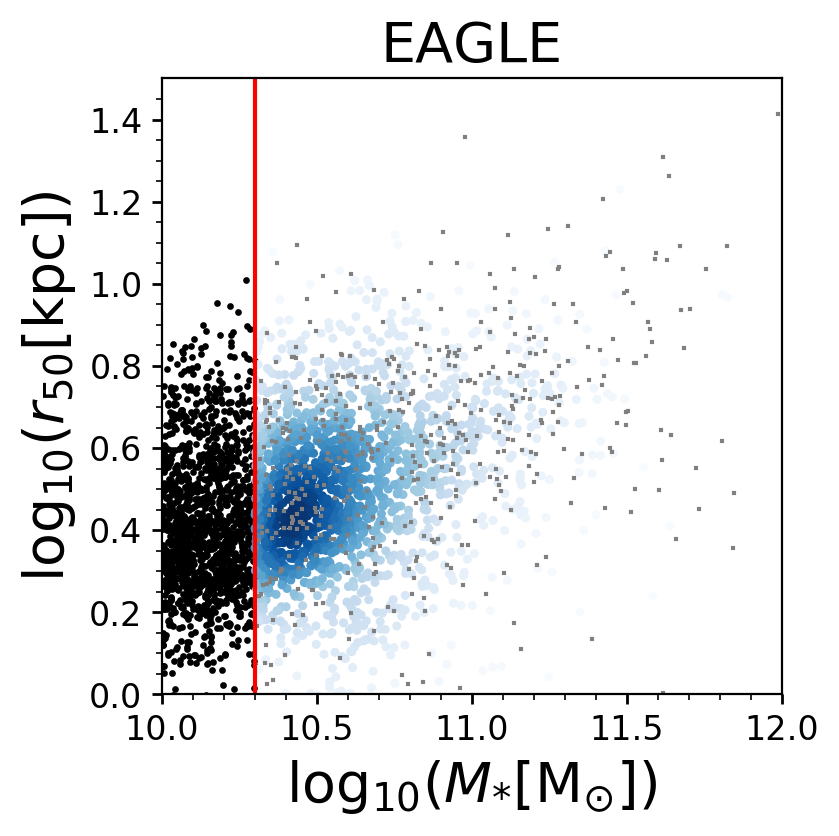}\centering\includegraphics[width=5.6cm]{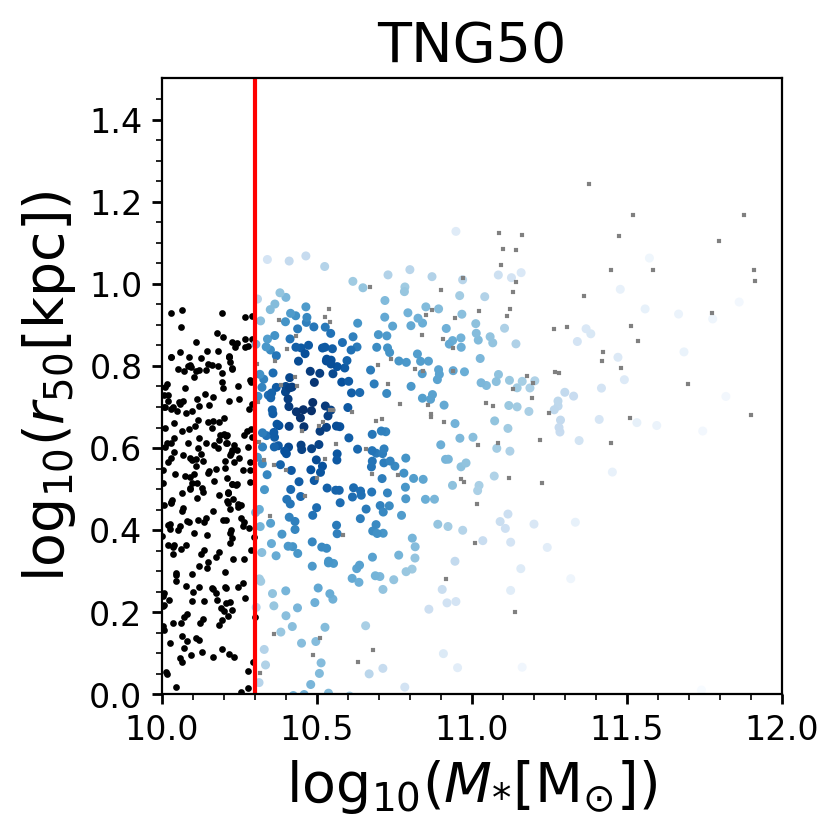}
\caption{Galaxy stellar mass-size relation of simulations analysed in this paper, from left to right are TNG100, EAGLE and TNG50, respectively. We choose the stellar mass defined within a spherical radius of 30 kpc, $M_{*}$, versus $r_{\rm 50}$ defined from 2D image. The galaxies kept in our analysis with $M_{*}>10^{10.3}M_{\odot}$ are coloured blue, the galaxies with on-going mergers are removed from our analysis and coloured grey.}
\label{sample selection}
\end{figure*}
%-------------------------------------------------------------------

\section{Introduction}
Galaxies grow by notably two channels: in situ star formation and ex situ galaxy-galaxy mergers, which lead to formation of different galaxy structures.
Galaxy structures, i.e. disk, bulge, and bar identified in the Hubble diagram \citep{1926ApJ....64..321H}, are taken as fossil records of galaxy assembly histories. It is generally believed that disks are formed by in situ star formation from regularly rotating gaseous disks \citep{1980MNRAS.193..189F}; classical bulges are formed by violent processes like protogalactic collapse or galaxy-galaxy mergers \citep{1977egsp.conf..401T}; and bar or pseudobulges are results of secular evolution such as disk instability \citep{2004ARA&A..42..603K}. Major mergers are thought to also play a significant role in the formation of stellar halos by destructing stellar disks \citep{2010MNRAS.406.2405B, 2011MNRAS.416.1654B, 2015ApJ...799..184P}; and minor mergers can cause significant growth of both the stellar and dark matter halos, resulting in a significant increase of the galaxy size \citep{2012MNRAS.425.3119H}. Galaxy structures are expected to provide valuable insight into their merger histories.

%\LZ{please complete the citing of references that I added. Most of them are from Remus&Duncan2022} \textbf{finish}

%Different regions of the galaxy represent distinct formation mechanisms. Some studies try to quantify the relationship between structure and ex situ. In-situ stars are thought to be formed through star-forming mechanisms within galaxies and ex situ stars are formed through merging events \citep{2010ApJ...725.2312O}. 
A lot of effort has been made in the literature trying to infer the merger history of galaxies from their morphological structures. However, no promising morphological parameters have been found.
Semi-analytical simulations suggest that early type galaxies can be generally fitted by a double Sersic profile, and there is a transition radius from in situ dominated regions to ex situ dominated and from the inner steep to the outer shallower component \citep{2013MNRAS.434.3348C,2015MNRAS.451.2703C}. There is also such a transition from in situ dominated to ex situ dominated in Illustris simulations \citep{2016MNRAS.458.2371R}. 
This transition radius or the mass of the outer shallower component has been used as an indicator of the ex situ stellar mass of galaxies\citep{2018MNRAS.479.4760F, 2014MNRAS.443.1433D, 2013ApJ...766...47H, 2017A&A...603A..38S, 2020A&A...639A..14S}. However, ex situ stars are also found to contribute in the very inner regions of galaxies \citep{2022A&A...660A..20Z,2022ApJ...935...37R}, and can be dominant at all radii for the most massive galaxies \citep{2019MNRAS.487.5416T}. In particular, with the hydrodynamical cosmological simulation Magneticum, although a large fraction of galaxies still show a transition radius in the surface brightness profile, it does not correspond to the transition from in situ to ex situ dominant, and it is not obviously correlated with the ex situ mass of galaxies \citep{2022ApJ...935...37R}.

Although ex situ stars can be distributed throughout the galaxy, the total ex situ stellar mass is still an important parameter as a first-order description of the merger history of a galaxy. Recent works use machine learning tools like cINN and Random forest \citep{2001MachL..45....5B} to study merger history of galaxies, by combing various parameters including stellar mass, stellar populations, and a few morphological parameters: galaxy radius ($r_{\rm 90}$, $r_{\rm 50}$), concentration, luminosity fraction of disk, etc. These models predicate ex situ stellar mass fraction well with a typical scatter of $\sim 0.1$ compared to the ground truth with enough input information \citep{2023MNRAS.519.2199E, 2022MNRAS.515.3938S}. Similar model predictions are made using the 2D spatially resolved maps as model input, including the stellar kinematics, stellar age, and metallicity maps, as can be obtained from IFU observations \citep{2023MNRAS.523.5408A,2024NatAs...8.1310A}. However, none of the above works finds any morphological parameters important, more important than the total brightness, in predicting the ex situ stellar mass fraction; the radius $r_{\rm 90}$ measured from 3D is found to be important, but considering the observational uncertainty, $r_{\rm 90}$ measured from mock images is still less important \citep{2022MNRAS.515.3938S}.

The merger history of the MW and a few nearby galaxies has been quantitatively uncovered through their 3D chemodynamical structures. An ancient massive merger that occurred in the MW about 10 Gyrs ago was discovered by the Gaia-Enceledus-sausage structure \citep{2018Natur.563...85H,2018MNRAS.478..611B}. Multiple merger events have been further quantified by the orbits and chemical properties of stellar populations \citep{2020MNRAS.493.5195D} and globular clusters \citep{10.1093/mnras/staa2452} in the MW halo.

However, we cannot resolve the stellar motion and chemical properties of single stars or GCs in most of the nearby galaxies. In the past two decades, integral field unit (IFU) spectroscopic instruments have mapped thousands of galaxies across a wide range of masses and Hubble types \citep{10.1046/j.1365-8711.2001.04612.x,2012A&A...538A...8S,de4a22341d89431bbc440c3a1e1b8cc4,2011MNRAS.414..888E,2012MNRAS.421..872C,2015ApJ...798....7B}. In principle, information regarding stellar motions and chemical distributions of an external galaxy is included in these IFU data, though all blended along the line of sight.

Based on the IFU data, two independent methods have been developed to uncover the galaxy merger history: (1) one tries to constrain the global ex situ fractions \citep{2021MNRAS.502.2296D,2021MNRAS.507.3089D, 2020MNRAS.491..823B} or the mass of satellite mergers \citep{2019A&A...625A..95P,2019A&A...623A..19P, 2021MNRAS.508.2458M} based on the age and metallicity distributions of stars in the inner regions of galaxies obtained from IFU data by full-spectral fitting. This method can identify minor mergers because their accreted stars have a lower metallicity than in situ stars, but they become insensitive to major mergers with overlapping metallicities. (2) another method is to uncover the internal 3D chemo-dynamical structure of a nearby galaxy by creating population orbit superposition models \citep{2020MNRAS.496.1579Z}, and to use a dynamically defined hot inner stellar halo as an indicator of merger mass. The hot inner stellar halo, defined by stars on highly radially orbits similar to the MW Gaia-Enceledus-sausage structure, is found to be highly correlated with the total ex situ stellar mass and the most massive merger mass the galaxies have ever experienced \citep{2022A&A...660A..20Z}. The merger time can be quantified by comparing the stellar age distribution of the disk and other components considering the interaction of the disk and the halo \citep{2022A&A...664A.115Z}, also in a way comparable to the MW \citep{2018MNRAS.478..611B}. The merger mass and the merger time for NGC 1380 and NGC 1427 have been determined using this method \citep{2022A&A...664A.115Z}. 
Both methods require high-quality IFU data and expensive spectra fitting or dynamical models, thus only being applied to a few case studies.

 Inspired by these studies inferring galaxy merger history from the chemo-dynamical structures, we want to find morphological structures that mimic the dynamically defined hot inner stellar halo and thus can be applied to a large sample of galaxies from photometric surveys. Using mock images created from the cosmological simulation IllustrisTNG and EAGLE, we will investigate if we can find morphological structures that efficiently trace the galaxy merger history, and we will further create models using random forest to predict the ex situ stellar mass of galaxies by only using information obtained from photometric data.
 This paper is structured as follows. In Section \ref{2} we describe the mock images and the definition of the morphological parameters extracted from the mock images. In Section \ref{3}, we introduce the Random Forest method. In Section \ref{4} we create models that predict $f_{\rm exsitu}$, and show the influence of the importance of different parameters. In Section \ref{5}, we discuss the dependence on the quality of photometric images and on different simulations. We conclude in Section \ref{6}.

%---------------------------------------------------------------------
\section{Data}
\label{2}
%---------------------------------------------------

%---------------------------------------------------
\subsection{Cosmological galaxy simulations}
%---------------------------------------------------
\begin{figure}
\subfloat{
		\includegraphics[scale=0.46]{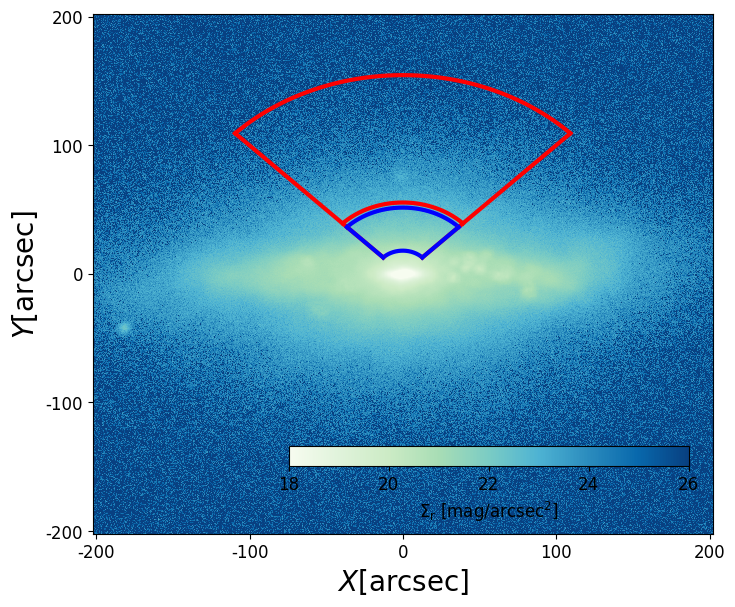}}
\quad        
\subfloat{
		\includegraphics[scale=0.4]{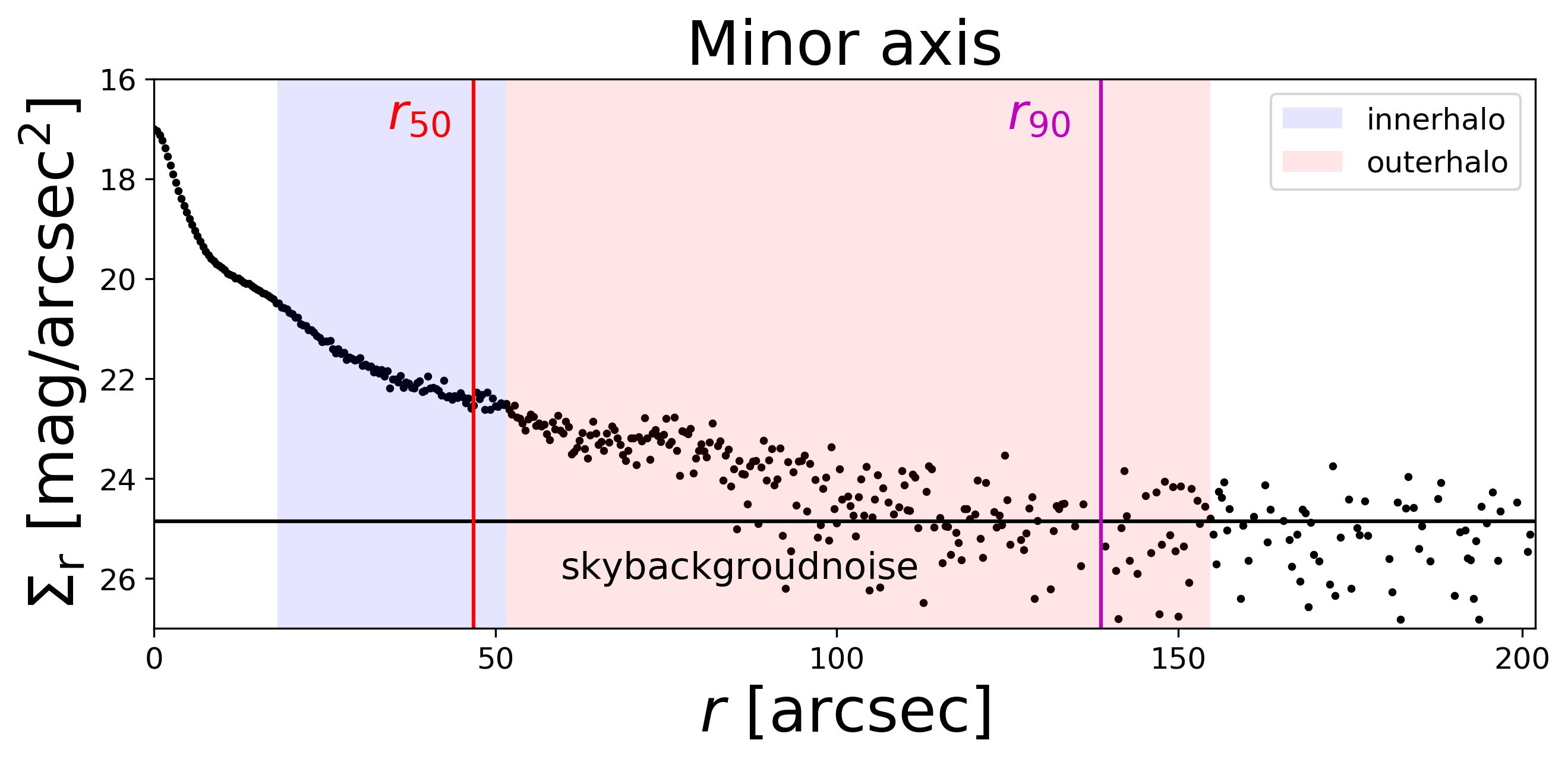}}
\caption{A SDSS-like $r$-band image created from a TNG100 galaxy subhalo 6 at z=0 projected near edge-on, and placing it at the distance of 40 Mpc. {\bf Top:} 2D image. The sector enclosed by blue is defined as the inner halo (3.5 kpc-10 kpc) and that in red is the outer halo (10 kpc-30 kpc). {\bf Bottom:} surface brightness profile along the minor axis. The black horizontal line indicates the background noise of the sky $\Sigma_{r,0}$, the red and magenta vertical lines mark $r_{\rm 50}$ and $r_{\rm 90}$ obtained from the petrosian radius. 
}

\label{TNG100SDSSlikeID6}
\end{figure}
Cosmological hydrodynamical simulations for the formation and evolution of galaxies have successfully reproduced galaxies in relatively large cosmic volumes. 
 IllustrisTNG simulations \citep{2018MNRAS.473.4077P,2018MNRAS.480.5113M,2019MNRAS.490.3234N,2018MNRAS.477.1206N,2018MNRAS.475..676S} have been successful in reproducing a wide range of observational findings \citep{2019ComAC...6....2N}. These include the galaxy mass-size relation at $0 < z < 2$ \citep{2018MNRAS.474.3976G}, but also the gaseous and stellar disk sizes and heights \citep{2019MNRAS.490.3196P}, galaxy colors, the stellar age and metallicity trends at z$\sim$0 as a function of galaxy stellar mass in comparison to SDSS results \citep{2018MNRAS.475..624N}, and resolved star formation in star-forming galaxies \citep{2021MNRAS.508..219N}, as well as the characteristics of the stellar orbit distributions from the CALIFA survey \citep{2019MNRAS.489..842X} and the kinematics of early-type galaxies in comparison to data from ATLAS-3D, MaNGA, and SAMI \citep{2020A&A...641A..60P}. It comprises three flagship runs: TNG300, TNG100 and TNG50 with different cosmological volumes and stellar particle resolutions (see Table~\ref{tab:sim_info}).  

The EAGLE \citep{2015MNRAS.446..521S,2015MNRAS.450.1937C} simulations have also been shown to successfully reproduce a range of observations of galactic properties, including the galaxy stellar mass function, the Tully–Fisher relation, and the galaxy mass-size relation. The galaxy sizes as a function of stellar mass generally agree with the SDSS results \citep{2015MNRAS.446..521S}. Galactic structures, such as disks and bulges, are well resolved, and the Hubble sequence is in place \citep{2016MNRAS.462.1470L,2019MNRAS.483..744T}. 

In this work, we use the publicly available data of TNG100 and TNG50 \citep{2019ComAC...6....2N} and the fiducial EAGLE simulation \citep{2016A&C....15...72M}. TNG100 and EAGLE are used for most of the analysis; they have similar cosmological volumes and stellar particle resolutions but are produced by substantially different numerical codes and with different galaxy formation models. The cosmological volumes are large enough that we can have enough galaxies for our analysis, and the spatial resolution is still high to allow us to investigate the galaxy structures at sub-kpc scale. 

There are 4133 galaxies in TNG100 and 2137 galaxies in EAGLE with $M_{*}>\rm 10^{10.3}$\,\Msun\, at z=0. We eliminated galaxies with ongoing mergers adopted by eyes, which are 673 and 458 in TNG100 and EAGLE, respectively; a few of such galaxies are illustrated in the Appendix Fig.\ref{fig:ongoing_merger}. 
There are also some galaxies with an obviously wrongly defined ex situ fraction by misidentification of its main progenitor galaxy. 
We identified 102 and 72 such galaxies in TNG100 and EAGLE from their merger tree and also excluded them from our sample. In the end, we have 3377 and 1620 galaxies from TNG100 and EAGLE, respectively. TNG50 has higher resolution compared to TNG100, but with a smaller cosmological volume. With similar selection, we have 443 galaxies from TNG50. The basic information of these three simulations is shown in Fig. \ref {sample selection} and is listed in Table~\ref{tab:sim_info}.

\begin{table*}
\def\arraystretch{1.5}
\caption{Basic information regarding the publicly available cosmological simulations we use or refer to in this paper. 
} 
\scriptsize\centering
\label{tab:sim_info}
\begin{tabular}{*{6}{l}}
\hline
       Name  & softening length[kpc] & $m_{\rm baryon}$ [$M_{\odot}$]  & $m_{\rm DM}$ [$M_{\odot}$] & $L_{\rm box}$ [cMpc] & NO of galaxies ($M_*>10^{10.3}$)\\
       \hline
TNG50 &0.3 &  8.5e4 & 4.5e5 & 50 & 443\\
TNG100 &0.7 &  1.4e6 & 7.5e6 & 100& 3377\\
EAGLE  &0.7 &  1.8e6 & 9.7e6 & 100& 1620\\

\hline
\hline
\end{tabular}
\tablefoot{From left to right, the columns show the name of the simulation, softening length, mass of stellar particles or gas cells, mass of dark matter (DM) particles, the side length of the simulation box and the number of galaxies we use in the paper. }
\end{table*}

%---------------------------------------------------
\subsection{Mock images}
\label{2.2}
%---------------------------------------------------
We created mock photometric images from TNG100 and EAGLE galaxies to mimic the SDSS and HSC observations for nearby galaxies. 
We take a few steps to create the mock images:
First, we read the coordinates and absolute magnitude of the stellar particles for each simulated galaxy and aligned the three main axes with $x$, $y$, and $z$. Second, we smooth the particles employing the spline kernel that is commonly used to Smooth the Particles in Hydrodynamical simulations \citep{1985A&A...149..135M,1992ARA&A..30..543M}. Third, we project these galaxies onto the sky plane near edge-on with inclination angles between 80 and 90 degrees. Fourth, we place the galaxy at a certain distance and divide it into pixels with certain pixel size on the 2D sky plane. Both TNG and EAGLE provide the particle luminosity in a few photometric bands including $r$, $g$, etc. We add the luminosity contribution of smoothed particles along the line of sight to construct the light in each pixel. Finally, we convolve the image with a Point-Spread-Function (PSF) kernel with a certain Full Width at Half Maximum (FWHM) using filter2D from OpenCV\footnote{\url{https://docs.opencv.org/3.4/d4/dbd/tutorial_filter_2d.html}} and add sky background noise $\Sigma_{0}$. 

We do not include more complex effects, such as gas or dust extinction, in our mock images. Our analysis in the following will show that we mainly rely on parameters related to the stellar halo where gas and dust should not play an important role \citep{2019MNRAS.483.4140R}.

All galaxies are projected nearly edge-on, and we created several versions of mock data by mimicking SDSS or HSC observations, and by placing the galaxies at different distances, for each galaxy we create images in the $r$ and $g$ bands. For SDSS-like images, we use a pixel size of 0.396 arcsec, a PSF kernel with FWHM of 1.32 arcsec, and sky background noise $\Sigma_0$ of 26.86 mag/pixel in the $r$ band, and 27.40 mag/pixel in the $g$ band\footnote{\url{https://www.sdss4.org/dr17/imaging/other_info/}}, following the typical quality of images in SDSS data release 17 \citep{2022ApJS..259...35A}. For HSC-like images, we use a pixel size of 0.168 arcsec, a PSF kernel with FWHM of 0.75 arcsec and sky background noise $\Sigma_0$ of 32.5 mag/pixel in the $r$ band, and 32.13 mag/pixel in the $g$ band \footnote{\url{https://hsc-release.mtk.nao.ac.jp/doc/}}, which are typical quality of images in HSC Public Data Release 3 \citep{2022PASJ...74..247A}.

For SDSS-like observations, we created several versions of mock images by placing all galaxies at the distance of 40, 100, 200, 400, 600 Mpc, and HSC-like observations with 40, 200, 400, 600, 1000, and 1500 Mpc to investigate how the quality of images will affect the results. In addition, we created a version of clean images by placing the galaxies at 40 Mpc and not including PSF or background noise for galaxies in TNG100, EAGLE, as well as TNG50. Note that we only created the clean image for TNG50 galaxies because it is expensive to smooth the particles with large numbers. We will use clean images for TNG50 when cross-validating the model with TNG50 in the following analysis.

We show an SDSS-like image in the $r$ band that we created from the TNG100 subhalo ID6 placing it at a distance of 40 Mpc from us in Fig.\ref {TNG100SDSSlikeID6}. We will use SDSS-like images at 40 Mpc as default for most of the analysis throughout the paper.

%---------------------------------------------------
\subsection{Definition of parameters}
\label{2.3}
%---------------------------------------------------
We aim to uncover the ex situ stellar mass fraction of galaxies ($f_{\rm exsitu}$). 
For galaxies from cosmological simulations, we define the stellar mass of the galaxy $M_*$ as the mass of all particles within 30 kpc sphere 
 of the galaxy.
We identify ex situ particles that do not belong to the main progenitor branch but are accreted from other progenitors. Ex situ mass is defined as the mass accreted and still exists in the galaxy at z = 0. The definition means that we do not consider the mass accreted in the past but stripped by other galaxies in our ex situ mass
\citep{2016MNRAS.458.2371R,2017MNRAS.467.3083R,2019MNRAS.483.4140R,2018MNRAS.473.4077P}. The ex situ stellar mass fraction ($f_{\rm exsitu}$) is defined as the ratio of ex situ stellar mass to the total stellar mass of the galaxy at $z=0$.

We define a few parameters that can be directly measured from the mock photometric images and two parameters from the LOS velocity distribution of a single aperture mimic the single-fibre spectroscopic observation. 
The parameters we define for each galaxy are as follows:
\begin{enumerate}

\item The absolute magnitudes in the $r$ and $g$ bands, $M_{\rm r}$ and $M_{\rm g}$ are determined by the luminosity in the corresponding band of all particles within 30 kpc of the galaxy, which is approximately with the Petrosian aperture as \citep{2018MNRAS.475..624N} shows.

\item Galaxy colour $\emph{g-r}$ defined as $M_{\rm g} - M_{\rm r}$.

\item Galaxy radii $r_{\rm 50}$ and $r_{\rm 90}$ determined from the $r-$band mock image. These parameters encompass 50\% and 90\% of the flux within a mock image. The total flux refers to the Petrosian flux, which is calculated as that within twice the Petrosian radius \citep{2002AJ....123..485S}. The Petrosian radius is defined as the circular radius at which the local surface brightness decreases to 20\% of the mean surface brightness within the aperture. 
In the bottom panel of Figure \ref {TNG100SDSSlikeID6}, we mark $r_{\rm 50}$ and $r_{\rm 90}$ of the galaxy defined in this way.

\item Concentration defined as 
$C=5\times \log_{10}(r_{\rm 90}/r_{\rm 50})$ 
 \citep{2003ApJS..147....1C}.

\item Luminosity fraction of inner and outer halos, $f_{\rm innerhalo}$ and $f_{\rm outerhalo}$. The inner and outer halo are defined by a sector with $45-135$ degrees from the major axis of the disk, with distances ranging from $3.5-10$ kpc and $10-30$ kpc, respectively, as shown in Figure \ref {TNG100SDSSlikeID6}. We used $r$-band luminosity weighted $f_{\rm innerhalo}$ and $f_{\rm outerhalo}$ for most of the analysis, calculated as the ratio of $r$-band luminosity within the inner and outer halo regions to the total luminosity within 30 kpc. The inner and outer halos are defined to avoid contamination of the disk and the compact bulge component. The separation of the bulge and inner stellar halo at $r=3.5$ kpc follows the definition of a dynamically hot inner stellar halo in \citet{2022A&A...660A..20Z}. 

\item The inner and outer surface brightness gradients, $\nabla \rho_{\rm inner}$ and $\nabla \rho_{\rm outer}$, along the minor axis by the same sector used for defining halos. The inner gradient $\nabla \rho_{\rm inner}$ is defined from the centre to 10 kpc, $\nabla \rho_{\rm inner} = (\rho_{\rm 9-11 kpc} - \rho_{\rm 0-3 kpc})/(10-1.5)$, and the outer gradient $\nabla \rho_{\rm outer}$ is from 10 to 30 kpc, $\nabla \rho_{\rm outer} = (\rho_{\rm 29-31 kpc} - \rho_{\rm 9-11 kpc}) / (30-10) $, where $\rho_{\rm 0-3 kpc}$, $\rho_{\rm 9-11 kpc}$, and $\rho_{\rm 29-31 kpc}$ are taken the average of surface brightness within the sector and in radii ranges of 0-3 kpc, 9-11 kpc, and 29-31 kpc, respectively.

\item The inner and outer color gradients, $\nabla \rm (\emph{g-r})_{inner}$, $\nabla \rm (\emph{g-r})_{outer}$, along the minor axis, defined in the same regions as $\nabla \rho_{\rm inner}$ and $\nabla \rho_{\rm outer}$, but taken the color $\emph{g-r}$ rather than the surface brightness.

\item Velocity dispersion $\sigma_v$ and kurtosis $h_{4}$ from a single aperture spectroscopic observation. We take a single aperture with radius size of 3 arcsec at the galaxy center, and extract the line-of-sight velocity distribution from all the particles in this aperture, to mimic that which could be obtained from a single fiber spectroscopic observation of each galaxy. We fit the LOS velocity distribution by a Gaussian-Hermit function, and extract the velocity dispersion $\sigma_v$ and the Kurtosis $h_{\rm 4}$ from the Gaussian-Hermit fitting \citep{1993MNRAS.265..213G,1993ApJ...407..525V}.

\end{enumerate}

\begin{table*}
\def\arraystretch{1.5}
\caption{Observational parameters defined in this paper.
} 
\scriptsize\centering
\label{tab:para11}
\begin{tabular}{*{6}{l}}
\hline
\hline
Name & Symbol & Description\\

\hline
Photometric parameters extracted from images\\
Absolute magnitudes& $M_{\rm r}$ and $M_{\rm g}$ & $r$ and $g$ band luminosity in the within 30 kpc of the galaxy.\\

Colour & $\emph{g-r}$& $M_{\rm g} - M_{\rm r}$.\\

Radii& $r_{\rm 50}$ and $r_{\rm 90}$ & Radius of the galaxy containing 50\% and 90\% of the $r$-band luminosity.\\

Concentration& $C$& $C=5\times \log_{10}(r_{\rm 90}/r_{\rm 50})$ \\

Luminosity fraction of inner halo& $f_{\rm innerhalo}$ & Luminosity of inner halo 3.5-10 kpc.\\

Inner surface brightness gradient& $\nabla \rho_{\rm inner}$ &The luminosity gradient from 1.5 to 10 kpc\\

Inner color gradient & $\nabla (\emph{g-r})_{\rm inner}$& The \emph{g-r} gradient from 1.5 to 10 kpc\\

Luminosity fraction of outer halo& $f_{\rm outerhalo}$ & Luminosity of outer halo 10-30 kpc.\\

Outer surface brightness gradient& $\nabla \rho_{\rm outer}$ &The luminosity gradient from 10 to 30 kpc\\

Outer color gradient& $\nabla (\emph{g-r})_{\rm outer}$& The \emph{g-r} gradient from 10 to 30 kpc\\

\hline
Kinematical parameters from single-aperture spectrum \\
Velocity dispersion & $\sigma_v$&Extracted from the line-of-sight velocity distribution within 3 arcsec.\\
kurtosis & $h_{4}$&Extracted from the line-of-sight velocity distribution within 3 arcsec.\\

\hline

Sub0  &$M_r$, $r_{90}$, $f_{\rm innerhalo}$, $f_{\rm outerhalo}$, $\nabla \rho_{\rm outer}$ and $\nabla (\emph{g-r})_{\rm outer}$\\

Sub1 &$M_r$, $r_{90}$, $f_{\rm outerhalo}$, $\nabla \rho_{\rm outer}$ and $\nabla (\emph{g-r})_{\rm outer}$\\

\hline
\hline
\end{tabular}
\tablefoot{We evaluate the ability of all listed parameters in predicting $f_{\rm exsitu}$. We only use parameters in Sub0 and Sub1 for our final model construction.}
\end{table*}

\begin{figure*}
\centering
\flushleft\subfloat{
		\includegraphics[scale=0.45]{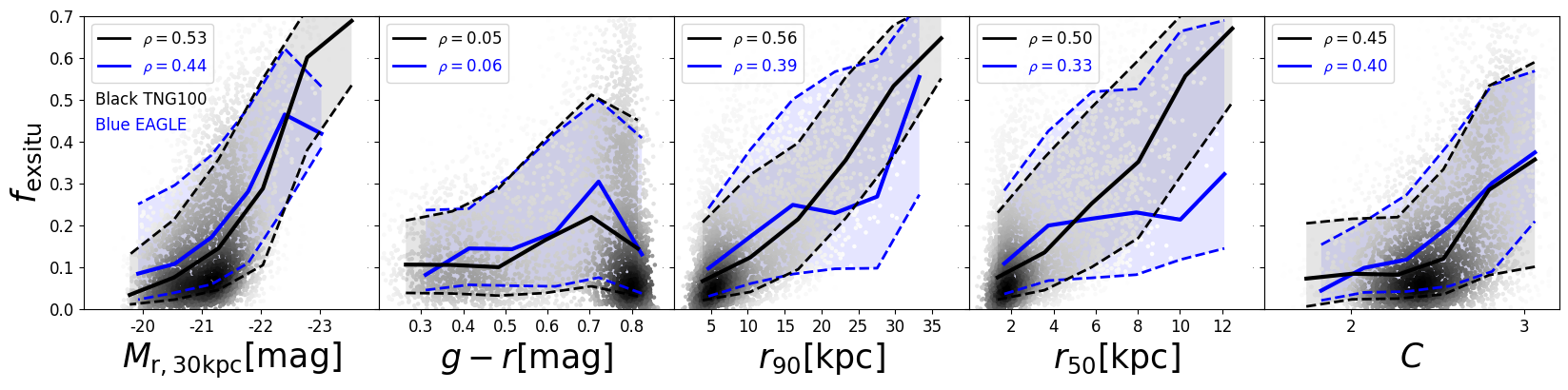}}
\flushleft\subfloat{
		\includegraphics[scale=0.45]{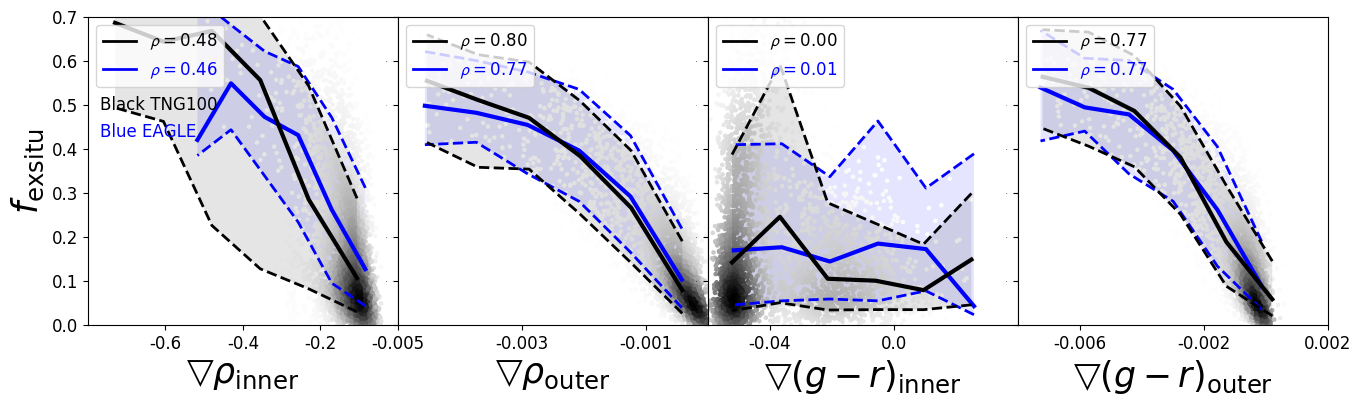}}
\flushleft\subfloat{
		\includegraphics[scale=0.45]{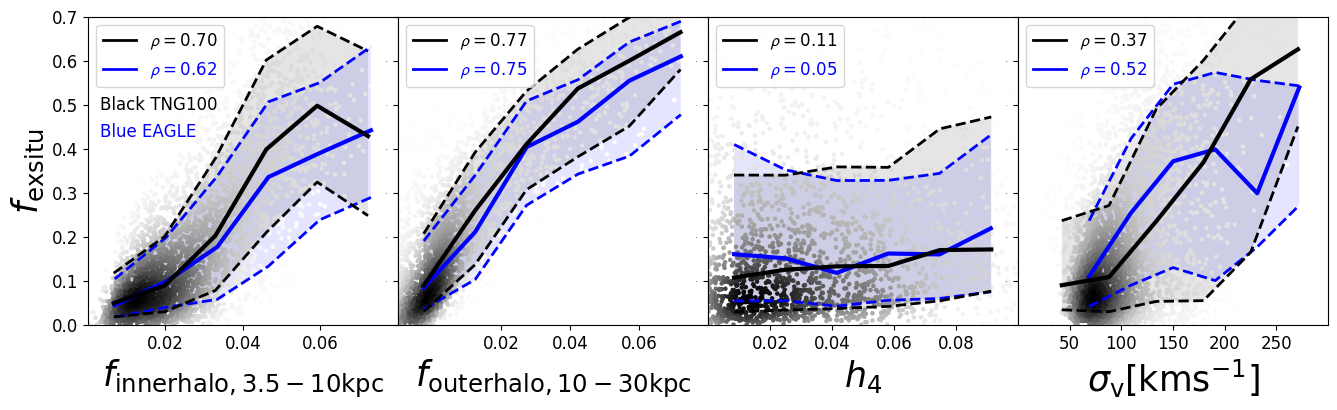}}
\caption{The correlations between the parameters extracted from mock SDSS photometric observations for galaxies at 40 Mpc and the ex situ stellar mass fraction of galaxies. The grey dots are TNG100 galaxies, the black solid and dashed curves are the running median and the $\pm 1 \sigma$ scatter of the TNG100 galaxies. The blue symbols for EAGLE galaxies. {The Spearman’s rank coefficient($\rho$)} of each correlation is labelled in the figure.  
}
\label{fig3:fexsitufeature}
\end{figure*}
In summary, we defined 12 parameters, $M_r$, $M_g$, $\emph{g-r}$, $r_{90}$, $r_{50}$, $C = 5\times \log_{10}(r_{90}/r_{50})$, $f_{\rm innerhalo}$, $f_{\rm outerhalo}$, $\nabla \rho_{\rm inner}$, $\nabla \rho_{\rm outer}$, $\nabla (\emph{g-r})_{\rm inner}$ and $\nabla (\emph{g-r})_{\rm outer}$ directly measured from the r and g band photometric images; and two parameters, $\sigma_v$ and $h_4$, from single aperture spectra. All the parameters defined are summarized in Table~\ref{tab:para11}. We properly included bias or uncertainties on the 12 parameters directly measured from the images by including observation effects in the mock images. For $\sigma_v$ and $h_4$, we did not consider bias or uncertainties caused by real observations.

%---------------------------------------------------
\section{Method}
\label{3}
%---------------------------------------------------
%---------------------------------------------------
\subsection{Decision tree}
%---------------------------------------------------
\begin{figure*}
\centering
\subfloat{\includegraphics[width=7cm]{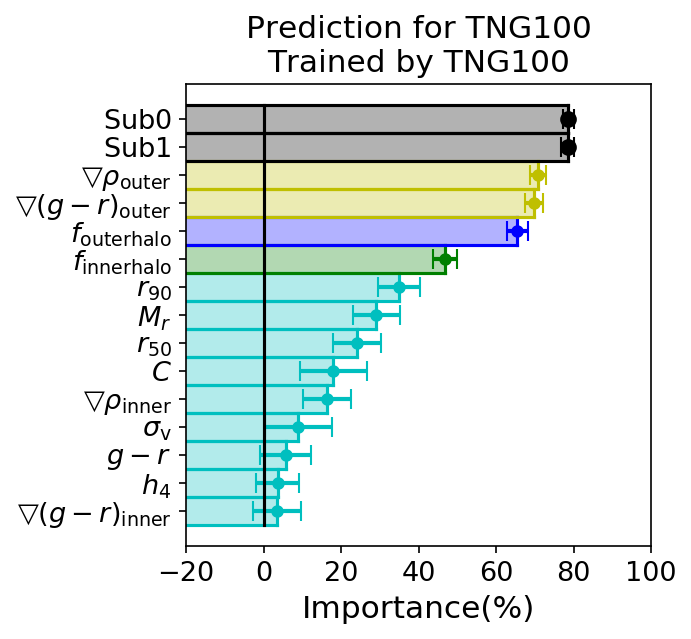}\includegraphics[width=7cm]{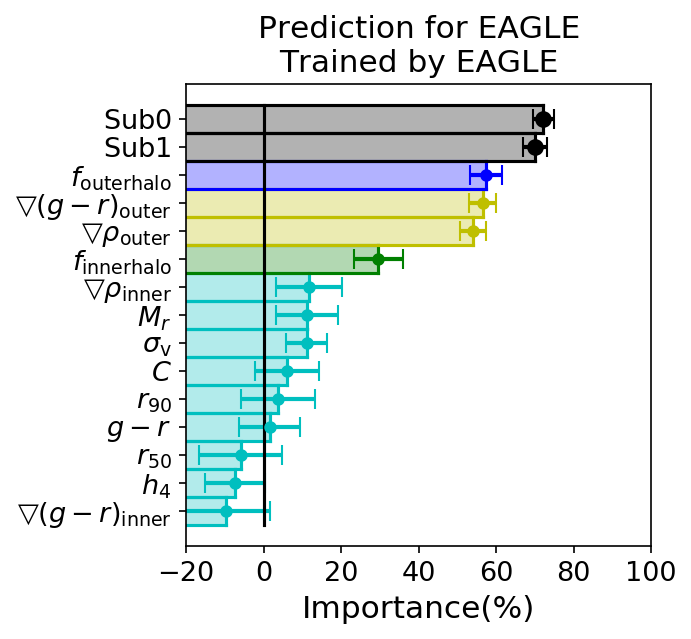}}
\quad
\subfloat{\includegraphics[width=7cm]{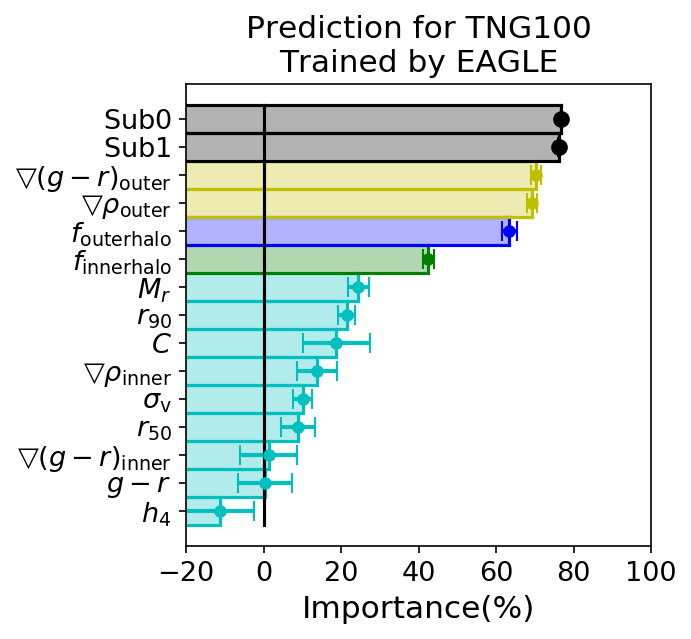}\includegraphics[width=7cm]{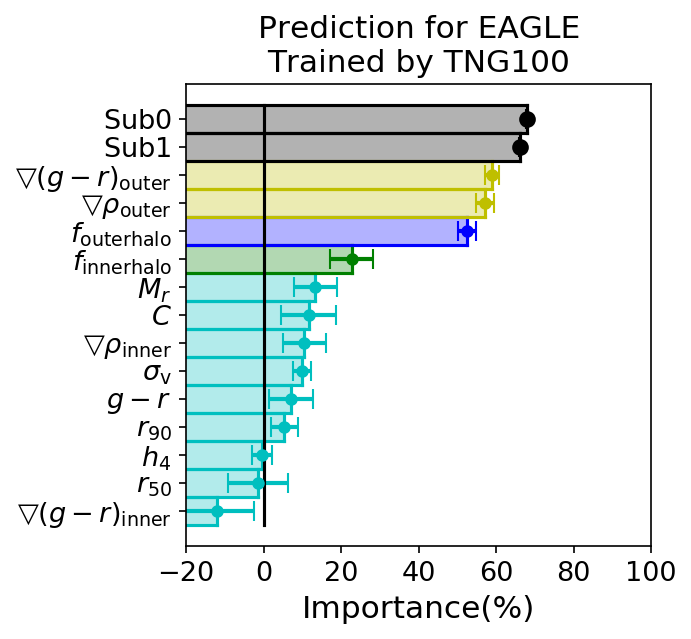}}
\caption{The importance ($r^2$) of the observational parameters in predicting the ex situ stellar mass fraction $f_{\rm exsitu}$.
The top panels are models trained and validated by TNG100 and EAGLE, respectively; the bottom panels are models cross-validated with TNG100 and EAGLE. In each panel, the two black points shows the importance of combined parameters. Sub0 including $\nabla \rho_{\rm outer}$, $\nabla (\emph{g-r})_{\rm outer}$, $f_{\rm outerhalo}$, $f_{\rm innerhalo}$, $M_{\rm r}$, $r_{\rm 90}$ and Sub1 including the same but not $f_{\rm innerhalo}$. The rest of the points show the importance of each single parameter as labeled. The error bar of the top two panels are the scatter of results from different training and validating sets and different hyperparameters of the RF models; the error bar of the bottom two panels are from different hyperparameters of the RF models.
}
\label{fig:R2_for_fex}
\end{figure*}

\begin{figure*}
\centering
\subfloat{\includegraphics[width=6.5cm]{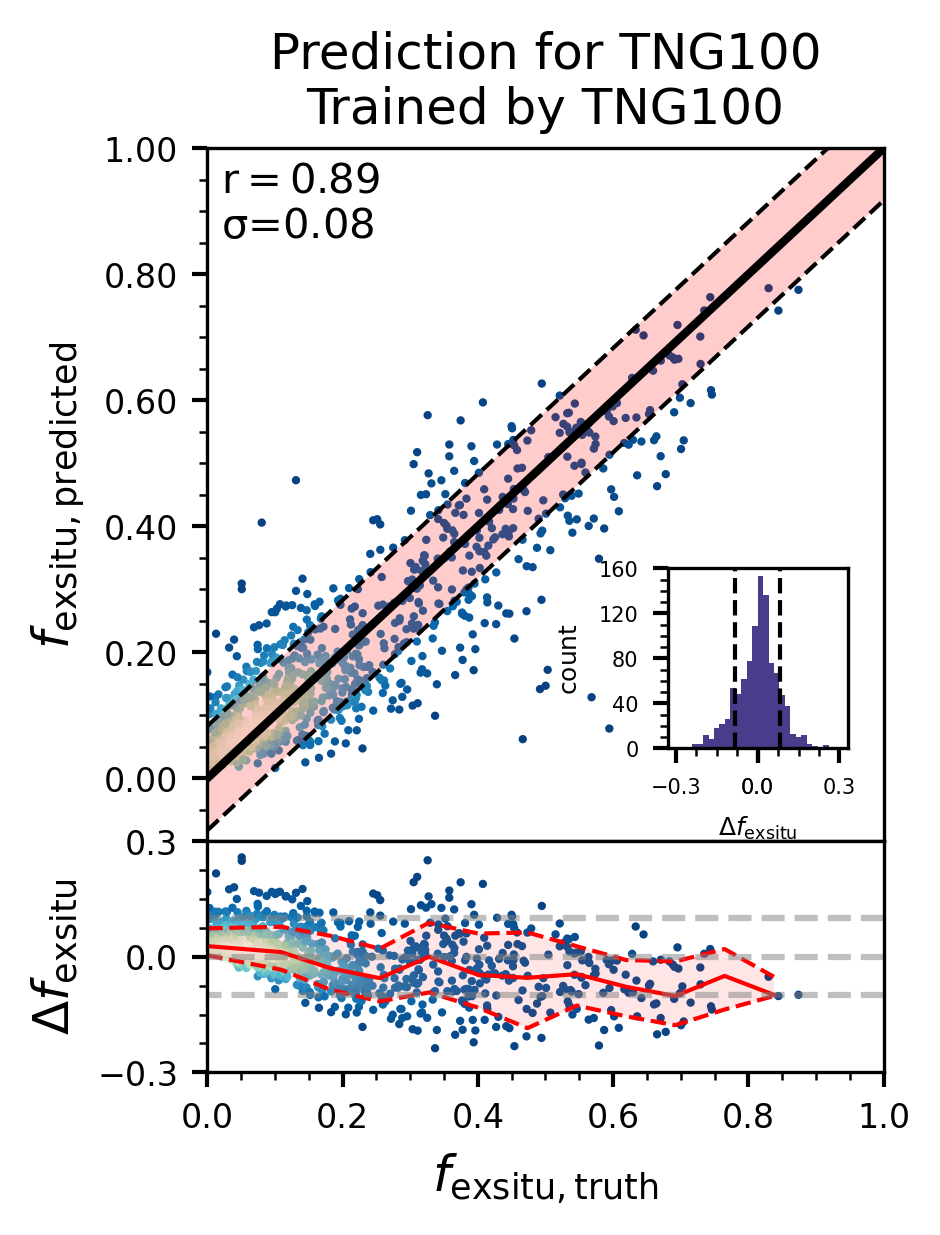}\includegraphics[width=6.5cm]{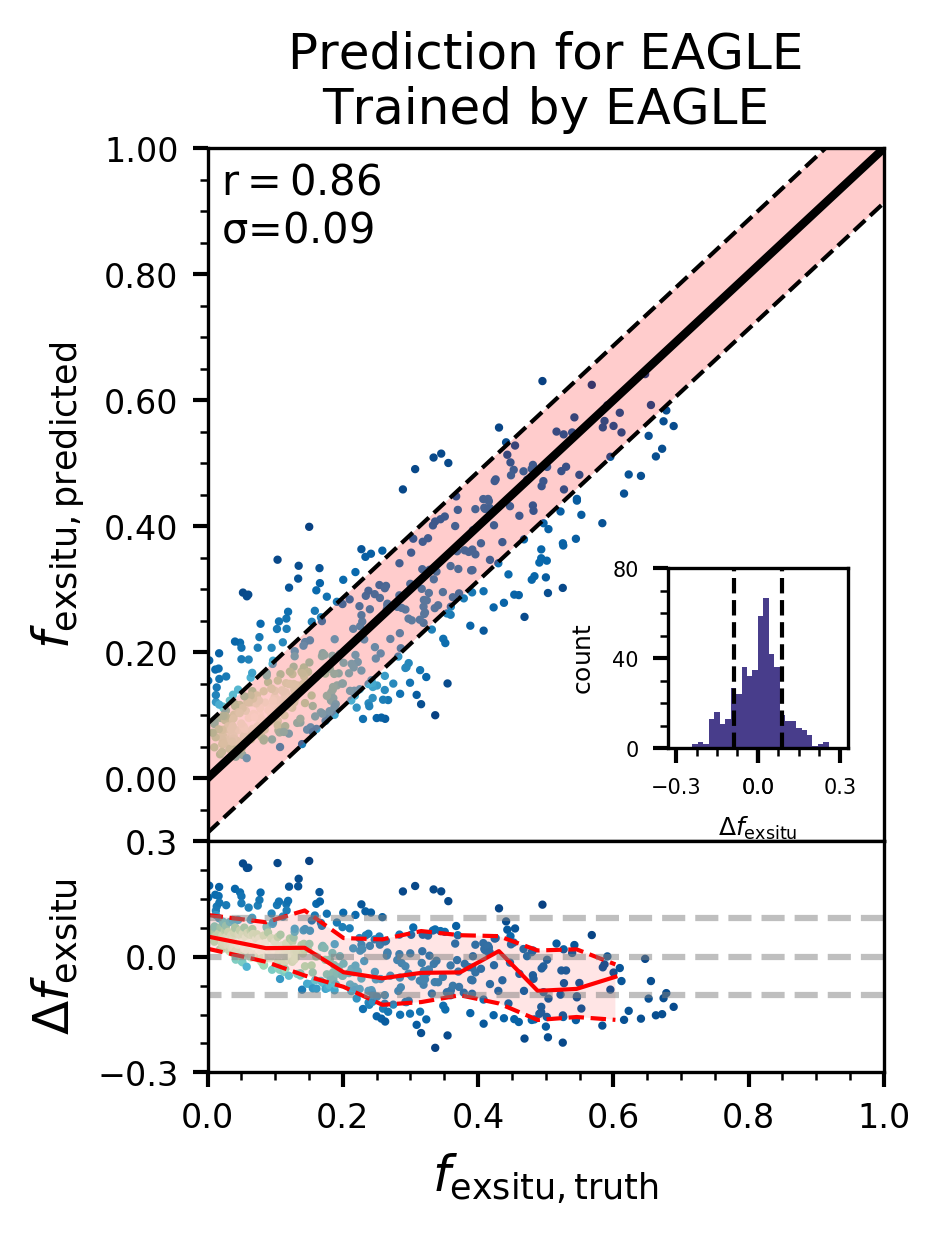}}
\quad
\subfloat{\includegraphics[width=6.5cm]{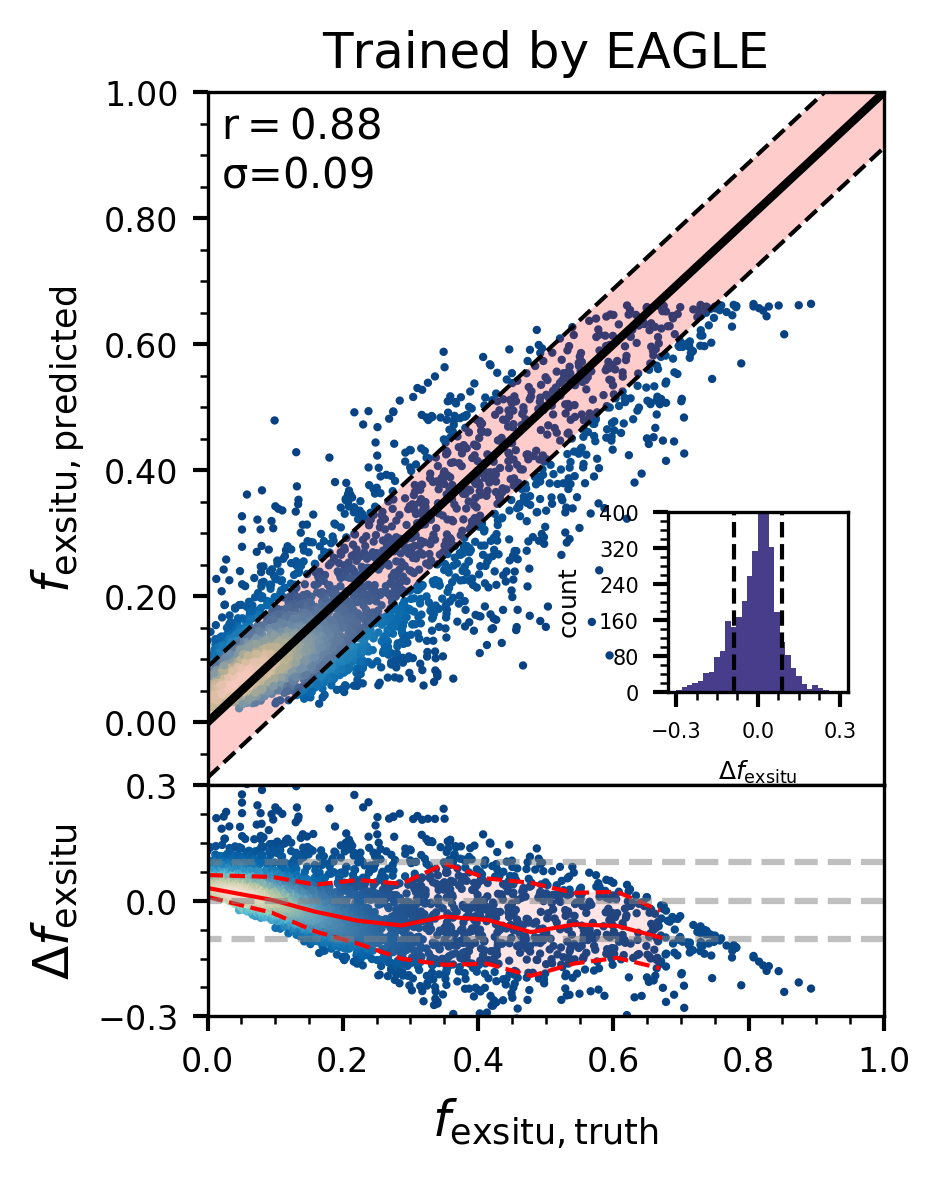}\includegraphics[width=6.5cm]{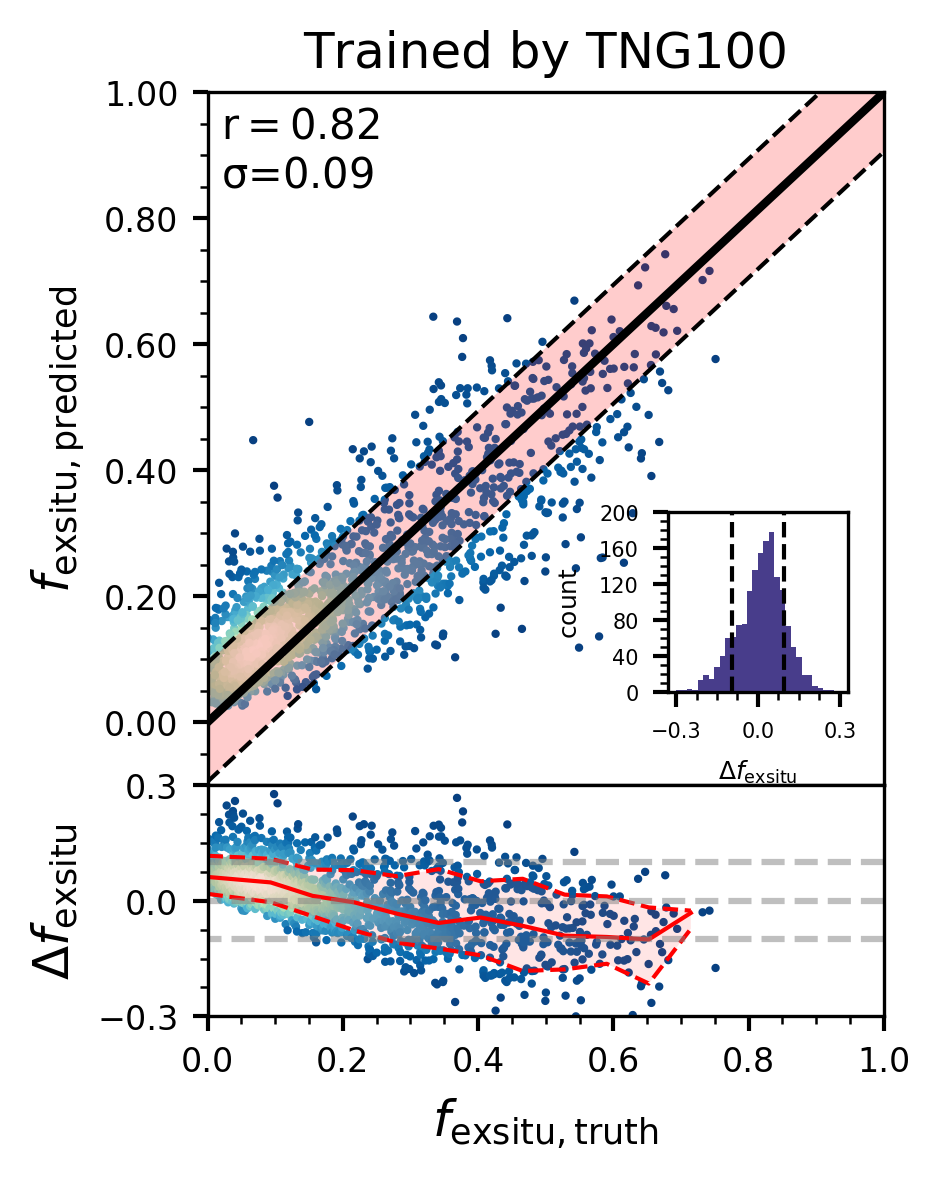}}
\caption{The model predicted $f_{\rm exsitu}$ vs. the ground truth. The different panels are models trained and validated by TNG100 and EAGLE in the top, and cross-validated with each other in the bottom, all with the combined parameters Sub0. For each column, we show the one-to-one comparison in the top panel: $r$ is $\sqrt{r^2}$, and $\sigma(\Delta f_{\rm exsitu})$ is the scatter of residual $\Delta f_{\rm exsitu}=f_{\rm exsitu,predicted}-f_{\rm exsitu,truth}$, the solid black line marks $y=x$, and the dashed black line represents ±1$\sigma(\Delta f_{\rm exsitu})$ scatter. The inset panel is the histogram of $\Delta f_{\rm exsitu}$. In the bottom panel, we show the residual $\Delta f_{\rm exsitu}$ as a function of galaxy $f_{\rm exsitu,truth}$: the red solid and dashed curves show the running median and ±1$\sigma(\Delta f_{\rm exsitu})$ scatter. }
\label{fig:fex_fextrue}
\end{figure*}

Before introducing Random Forest, let us briefly discuss the foundational concept of a decision tree. The decision tree \citep{Breiman1984} is a simple yet highly interpretable machine learning algorithm that aligns with human intuitive thinking, acting as a supervised learning algorithm based on the if-then-else rule. It establishes a mapping between properties and the value of an object in a tree-like structure. In this structure, each node represents an object, each forked path represents a possible property value, and each leaf node corresponds to the value of the object represented by the path from the root node to the leaf node.

%---------------------------------------------------
\subsection{Random Forest Method}
%---------------------------------------------------
The random forest \citep{2001MachL..45....5B} is an algorithm that uses the ensemble learning concept of bagging to combine multiple decision trees. It can be applied for clustering, classification, and regression analyses. For classifying an input sample, it is subjected to classification by each decision tree, and the classification results of these weak classifiers (i.e., decision trees) are aggregated to form a strong classifier (i.e., Random Forest). There are two key generation rules for each decision tree within the Random Forest algorithm. First, 
%when the size of the training set is N, each tree randomly selects N  training samples from the data set 
with data size $N$, if we set the training set size to $n$, then each tree randomly selects a training sample of $n$ (with $n<N$) from the data set using the bootstrap sample method, resulting in a training set that differs for each tree and contains repeated training samples. Second, of the $M$ features, $m$ features 
(with $m < M$) are randomly chosen from $M$ when splitting each node, and the best feature (with the maximum information gain) among these $m$ features is used for node splitting. Throughout forest growth, the value of $m$ remains consistent. The introduction of these two levels of randomness significantly influences the classification performance of Random Forest. Their incorporation helps prevent overfitting and enhances noise immunity. During the classification task, each decision tree classifies the newly entered samples, which ultimately contributes to the final output.

In this paper, we use the RandomForestRegressor class from the Python package scikit-learn \citep{scikit-learn} to construct the Random Forest. There are a few important hyperparameters that need to be considered for the Random Forest configuration: (1) the number of decision trees $n_{\rm estimators}$, for which we explore the optimal value in the range of 5 to 3000; (2) the number of features to consider at each node $n_{\rm max-features}$, given the total features of $M$, we adopt $n_{\rm max-features} = \sqrt{M}$; (3) the maximal depth allowed for each tree $n_{\rm max-depth}$, for which we search for the optimal value within the range of 10 to 500; (4) the minimal number of samples required for a node to be split $n_{\rm min-split}$, for which we seek the best value from the options of [2, 5, 8]; (5) the minimum number of samples required to form a leaf node $n_{\rm min-leafrange}$, which we explore the optimal value among [1, 2, 4, 8]. We allow a wide range for the two hyperparameters, $n_{\rm estimators}$ and $n_{\rm max-features}$, which are considered the most important hyperparameters in the RF method. 

To construct the model, we first divide our data sets into model and validation data sets by fixing the fraction of data in the two sets to be $7:3$. The model data set will be used to train and test the model; therefore, we further separate the model data set into training and testing data sets using the 3-fold cross-validation method of the GridSearchCV class in the sklearn package \citep{pedregosa2011scikit}. With the 3-fold cross-validation method, we randomly separate the model data set into three parts. We take two of the three parts to train the model, and the rest one to test the model each time, and repeat the processes three times to use the three parts as training and testing, in turn. We finally take the average results of the three models. This method ensures the robustness of the model and reduces the impact of data partitioning on the performance of the model. 

We then evaluate the model using the validating data and the results are quantified using the r-square ($r^2$) metric. 
\begin{equation}
\label{eq:r2}
     r^2 = 1 - \frac{\sum\limits_{i}(\hat{y_{i}}-y_{i})^2}{\sum\limits_{i}(\overline{y_{i}}-y_{i})^2},
\end{equation}
where $\hat{y_{i}}$ is the model prediction, $y_{i}$ is the ground truth, and $\overline{y_{i}}$ is mean of model predicted value. 

We also consider the uncertainty caused by partitioning the model and validation sets by performing the separation at random 50 times. For each separation, we create the model using the model data set and evaluate it using the corresponding validating data set. Finally, we average the $r^2$ value from the 50 models.

We understand that the machine learning models created from galaxy simulations may depend on the performance of simulations, including the resolution and physics in the galaxy formation.  
We create model A using $70\%$ of galaxies from TNG100 to train the model and the remaining $30\%$ galaxies to validate the model; we create model B in a similar way but using galaxies from EAGLE; for model C, we use EAGLE galaxies to train the model but TNG100 galaxies to validate the model; for model D, we use TNG100 galaxies to train the model but EAGLE galaxies to validate the model, as summarised in Table \ref{tab:model_info}. 
\begin{table}
\def\arraystretch{1.5}
\caption{The training and validating datasets used for the four models. 
} 
\scriptsize\centering
\label{tab:model_info}
\begin{tabular}{*{6}{l}}
\hline
       Model  & Training set & Validating set  \\
       \hline
A &70\%TNG100 galaxies&  30\%TNG100 galaxies\\
B &70\%EAGLE galaxies&  30\%EAGLE galaxies\\
C &All EAGLE galaxies&  All TNG100 galaxies\\
D &All TNG100 galaxies&  All EAGLE galaxies\\

\hline
\end{tabular}
\end{table}

%---------------------------------------------------
\section{Result}
\label{4}
%---------------------------------------------------

%----------------------------------------------------------
\subsection{Correlations between observations and $f_{\rm exsitu}$}
\label{2.4}
%----------------------------------------------------------

We have defined a series of parameters from the mock photometric observations, we used the SDSS-like mock data by placing galaxies at a distance of 40 Mpc for all the analysis in this section. In Figure \ref{fig3:fexsitufeature}, we show their correlations with the ex situ stellar mass fraction $f_{\rm exsitu}$ for TNG100 and EAGLE galaxies. The tightness of the correlations is quantified by Spearman rank coefficient of correlation ($\rho$), with $\rho=1$ indicating a perfect positive correlation; $\rho=-1$ indicating a perfect negative correlation; and $\rho=0$ indicating that there is no correlation. 

The outer surface brightness gradient $\nabla \rho_{\rm outer}$, the outer colour gradient of $\nabla (\emph{g-r})_{\rm outer}$, the luminosity fraction of the outer halo $f_{\rm outerhalo}$ and the inner halo $f_{\rm innerhalo}$ exhibit the strongest correlations with $f_{\rm exsitu}$ compared to other parameters. The magnitude of the galaxy $M_{\rm r}$, the radius of the galaxy $r_{\rm 90}$, $r_{\rm 50}$, the inner surface brightness gradient $\nabla \rho_{\rm inner}$, the concentration $C$, and the single aperture velocity dispersion $\sigma$ also exhibit moderate correlations with $f_{\rm exsitu}$. The galaxy colour $\emph{g-r}$, inner colour gradient $\nabla (\emph{g-r})_{\rm inner}$ and single aperture kurtosis $h_{\rm 4}$ have a weak correlation with $f_{\rm exsitu}$.

The correlations between the above observational parameters and $f_{\rm exsitu}$ are similar in TNG100 and EAGLE galaxies. Although the correlations are slightly stronger in TNG100 galaxies with higher $\rho$, there are no significant systematic offsets between TNG100 and EAGLE galaxies for the median curves of the correlations shown in Fig.~\ref{fig3:fexsitufeature}. Such correlations also exist and are similar in the TNG50 galaxies (see Figure \ref{fig3:appendix_tng50}). Thus, these correlations between the observational parameters we defined and $f_{\rm exsitu}$ are independent of the galaxy formation model and the simulation resolution, consistent with the relations found with the dynamically defined hot inner stellar halo \citep{2022A&A...660A..20Z}.

\begin{figure*}
\centering\subfloat{\includegraphics[width=7cm]{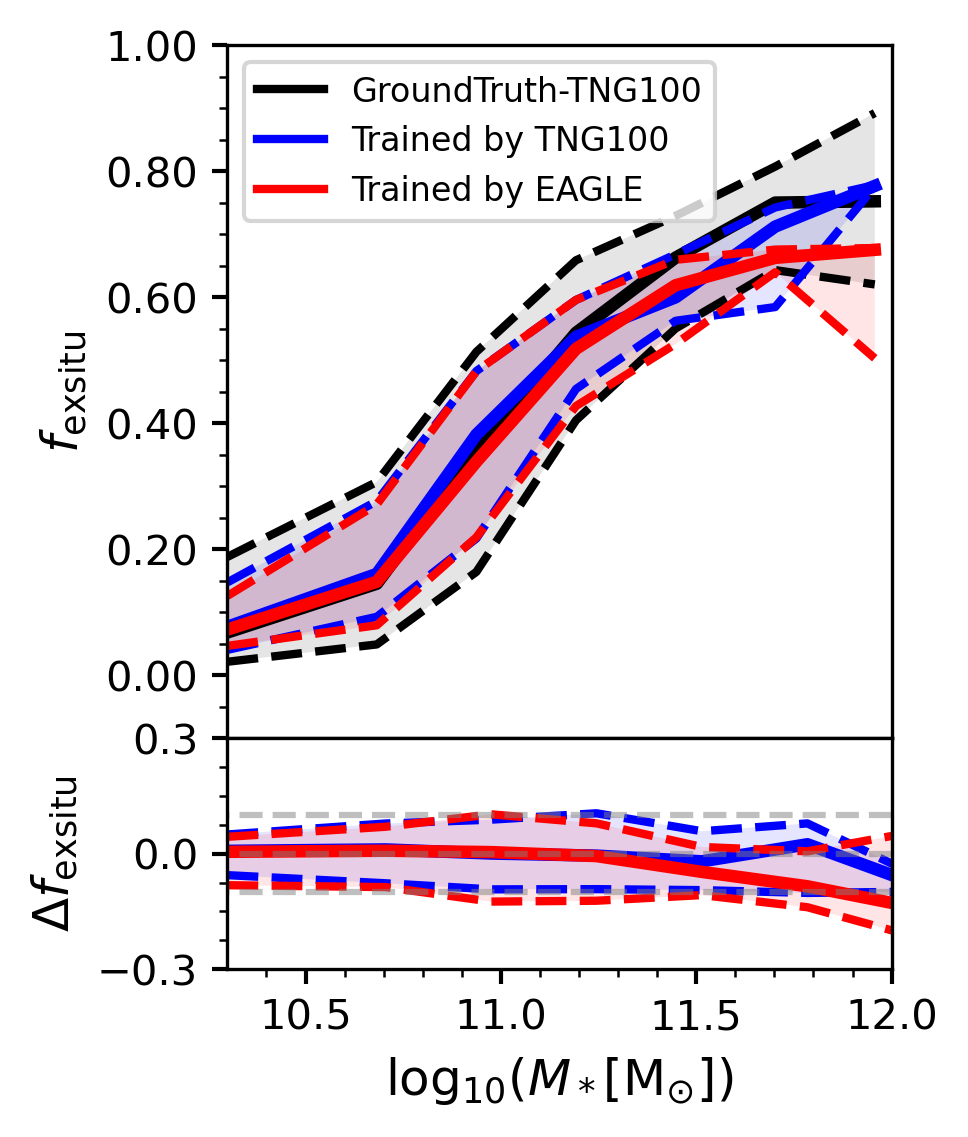}\includegraphics[width=7cm]{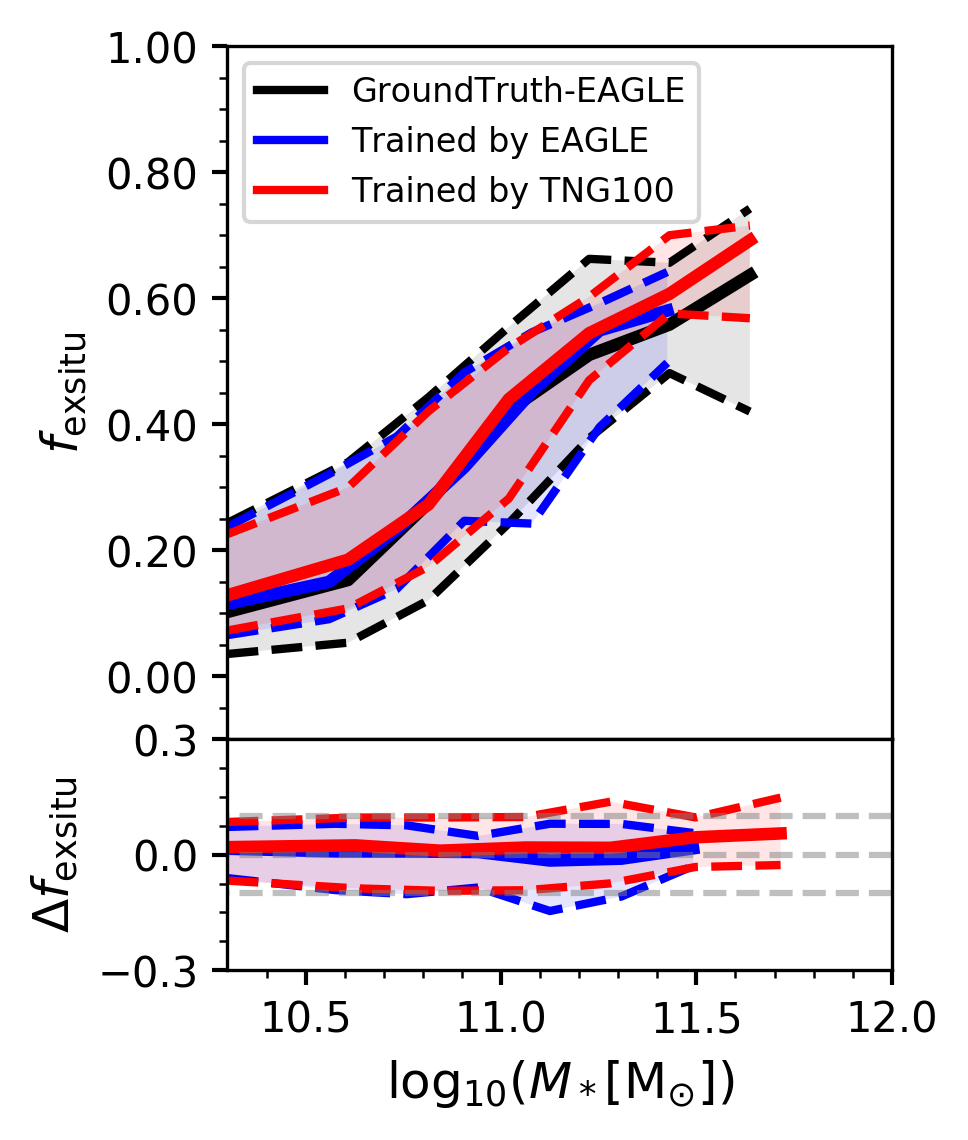}}
\caption{The $f_{\rm exsitu}$ as a function of stellar mass for model predications vs. ground truth. {\bf Left:} the black solid curve is the median of the ground truth for the TNG100 galaxies, the blue and red solid curves represent those predicted by Model A (trained by TNG100) and Model C (trained by EAGLE), the dashed curves are the $\pm 1 \sigma$ scatter. {\bf Right:} the black curves are the ground truth for EAGLE galaxies, the blue and red represent those predicted by Model B (trained by EAGLE) and Model D (trained by TNG100).}
\label{fig:fexvsmass}
\end{figure*}

%---------------------------------------------------
\subsection{The importance of the features}
\label{4.2}
%---------------------------------------------------
The careful choice of input features plays an important role in achieving the optimal performance of a machine learning model. We evaluated the importance ($r^2$, equation\ref{eq:r2}) of different parameters in predicting $f_{\rm exsitu}$ using RF. In order to check the dependence on the simulations, we evaluated the importance in four RF models: models trained and validated by TNG100 and EAGLE galaxies, respectively, and models cross-validated with TNG100 and EAGLE.

We show the importance of the parameters of the four models in Fig. \ref {fig:R2_for_fex}. In all four models, the outer surface brightness gradient $\nabla \rho_{\rm outer}$, the outer colour gradient $\nabla \rm{(g-r)}_{outer}$, the luminosity fraction of the outer halo $f_{\rm outerhalo}$ and the inner halo $f_{\rm innerhalo}$ are the four with the highest importance for the prediction of $f_{\rm exsitu}$, consistent with the strong correlations of these parameters with $f_{\rm exsitu}$ shown in Fig. \ref {fig3:fexsitufeature}.

The parameters of the next importance are different for models trained by TNG100 and EAGLE galaxies. In the model trained and validated by TNG100, the galaxy size $r_{\rm 90}$ is the next important feature, followed by the magnitude $M_r$ and the galaxy size $r_{\rm 50}$. In the rest of the models, $\nabla \rho_{\rm inner}$ or $M_r$ is the fifth important feature, and $r_{\rm 90}$ is not so important. 

The other parameters are not important in all models. 
We evaluated the importance of the combined parameters Sub0 that includes $\nabla \rho_{\rm outer}$, $\nabla \rm (\emph{g-r})_{outer}$, $f_{\rm outerhalo}$, $f_{\rm innerhalo}$, $M_{\rm r}$, $r_{\rm 90}$ and Sub1 including the same but not $f_{\rm innerhalo}$, the combination of parameters Sub0 has slightly higher or similar importance with Sub1 as shown in Fig. \ref{fig:R2_for_fex}. The other parameters are of little importance and do not help much in improving the model. In the following analysis, we take Sub0 as the default combination in the training of the models.

As shown in Fig. \ref{fig:R2_for_fex}, the uncertainties caused by the hyperparameters of the Random Forest model or by the division of the datasets are small. In the two lower panels, when the model sample is fixed at TNG100 and the validation sample is fixed at EAGLE, the scatter is entirely determined by the model error introduced by the hyperparameters. In this case, the spread of importance for the group of parameters, Sub0 and Sub1, is negligible.

%---------------------------------------------------------------
\subsection{Model predicted $f_{\rm exsitu}$ vs. ground truth}
\label{4.1}
%---------------------------------------------------------------

We train RF models using the combination of parameters Sub0 that includes $\nabla \rho_{\rm outer}$, $\nabla (\emph{g-r})_{\rm outer}$, $f_{\rm outerhalo}$, $f_{\rm innerhalo}$, $M_{\rm r}$, $r_{\rm 90}$. All these parameters are measured from photometric images. 
We still create four sets of models trained and validated by either TNG100 or EAGLE galaxies or cross-validated.

We show the performance of the model to predict $f_{\rm exsitu}$ in Fig. \ref{fig:fex_fextrue}. The top panel are models trained and validated by TNG100 and EAGLE galaxies, respectively, and the bottom panel are the models cross-validated with each other. The model trained and validated by TNG100 works well with $\mathcal{R}=0.89$ and the standard derivation of $\Delta f_{\rm exsitu} =f_{\rm exsitu,predicted}- f_{\rm exsitu,truth}$ to be 0.08. The model trained and validated by EAGLE works similarly well. In order to further check if the model can be transferred between different simulations, we validate the model trained by EAGLE with TNG100 galaxies, and vice versa. The models still work reasonably well in predicting $f_{\rm exsitu}$ for galaxies from a different simulation, although slightly worse than that from the same simulation. These findings align with the statistical consistency of the correlations found in TNG100 and EAGLE as shown in Fig. \ref {fig3:fexsitufeature}.

In all the models, there is a slight systematic bias that the models tend to overpredict $f_{\rm exsitu}$ for those galaxies with low $f_{\rm exsitu}$, and underpredict $f_{\rm exsitu}$ for the few galaxies with very high $f_{\rm exsitu}$. The standard derivation of $\Delta f_{\rm exsitu}$ is almost constant, with 0.1 across $f_{\rm exsitu}$. 

We further check how well the models reproduce $f_{\rm exsitu}$ as a function of galaxy stellar mass in Fig. \ref{fig:fexvsmass}. The models, either trained by TNG100 or EAGLE, reproduce well the $f_{\rm exsitu}$ as a function of stellar mass for both TNG100 and EAGLE galaxies, including the medium curve and the $1\sigma$ scatter. There is no systematic bias as a function of galaxy stellar mass, except for the most massive galaxies at $M_* \gtrsim 5\times 10^{11}$\,\Msun\,where we have very few galaxies in both the training and validation datasets.

\begin{figure*}
\centering\includegraphics[width=18cm]{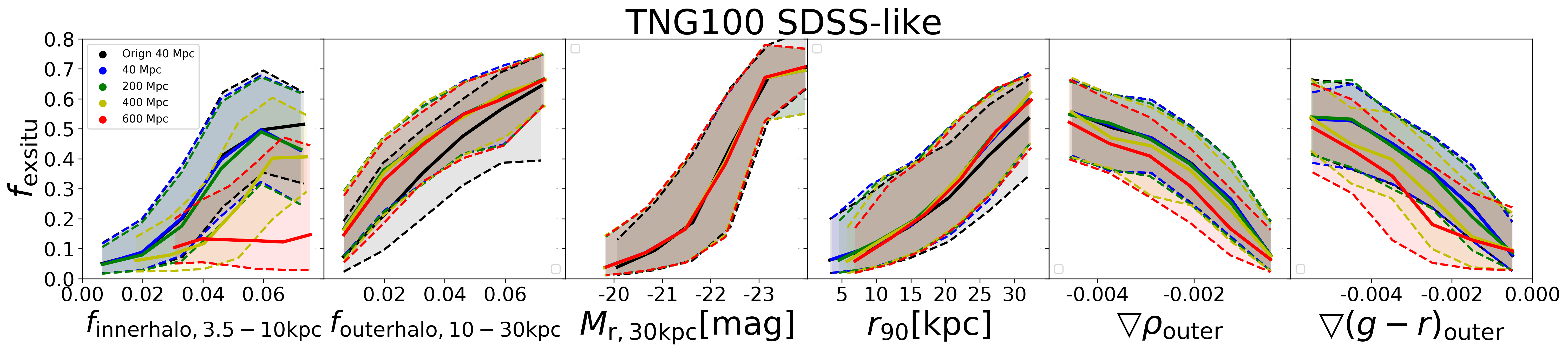}\centering
\caption{The parameters derived from mock images vs. $f_{\rm exsitu}$ for galaxies placed at different distances and with SDSS-like observations. In each panel, the black curves represent galaxies at 40 Mpc and without observational noise; the blue, green, yellow, red represent galaxies with observational noise and placed at 40, 200, 400 and 600, respectively. The solid curves are running median and the dashed curves are ±1$\sigma$ scatter. 
}
\label{SDSS distance}
\end{figure*}

\begin{figure*}
\centering\includegraphics[width=18cm]{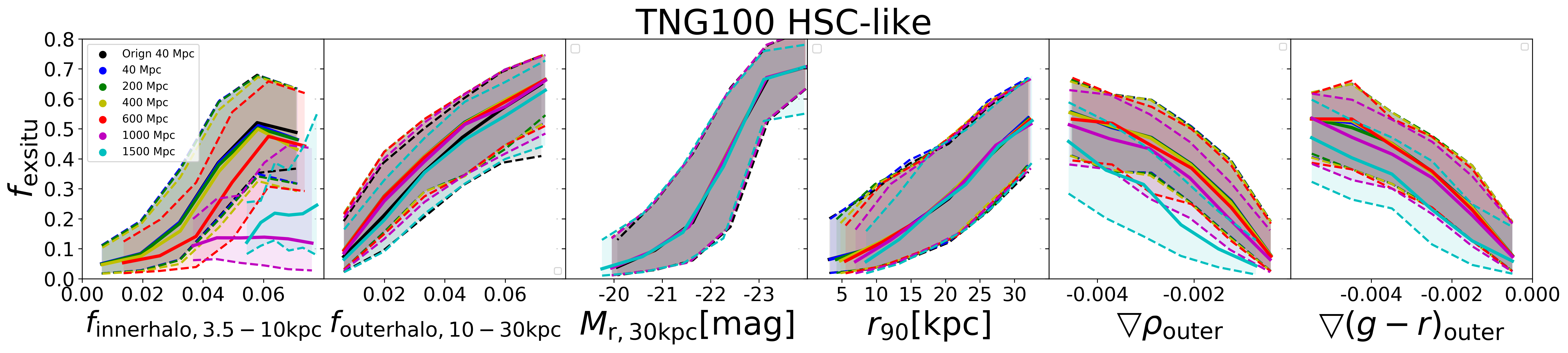}
\caption{Same as Fig.\ref{SDSS distance} but for galaxies with HSC-like observations. In each panel, the black curves represent galaxies at 40 Mpc and without observational noise, the blue, green, yellow, red, magenta and cyan represent galaxies with observational noise and placed at 40, 200, 400, 600, 1000, and 1500 Mpc, respectively.}
\label{HSC distance}
\end{figure*}

%---------------------------------------------------
\section{Discussion}
\label{5}
%---------------------------------------------------

%---------------------------------------------------
\subsection{The effects of observational noise}
\label{5.1}
%---------------------------------------------------
\begin{figure}
\centering\subfloat{\includegraphics[width=7cm]{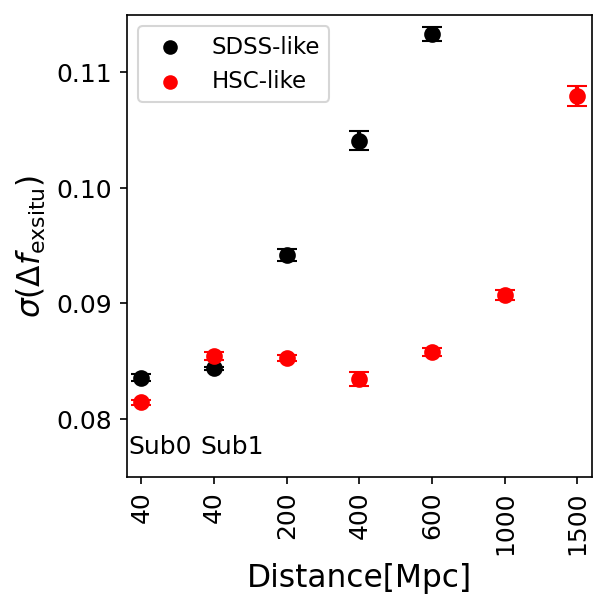}}
\caption{The standard deviation of model residuals $\sigma(\Delta f_{\rm exsitu})$, with a model trained with galaxies at 40 Mpc and tested with mock galaxies put at different distances as shown along the x-axis. Red points are HSC-like galaxies, and black points are SDSS-like galaxies.}
\label{fig:Sigmadistance}
\end{figure}
 We used mock images with SDSS-like observational data and placed galaxies at 40 Mpc in the above analysis. To understand how the results will be affected by the observational noise, here we create a few versions of mock images from TNG100 by placing the galaxies at 40, 200, 400, 600 Mpc with the SDSS-like observational noise and placing the galaxies at 40, 200, 400, 600, 1000, 1500 Mpc with HSC-like observational noise. In comparison, we also create a version of clean image placing at 40 Mpc and without observational noise.

We extract similarly parameters from these mock images and show the correlations between the photometric parameters and $f_{\rm exsitu}$ in Fig. \ref{SDSS distance} and Fig. \ref{HSC distance}. We focus on the six parameters that are of the highest importance in predicting $f_{\rm exsitu}$ here, i.e., the parameters in Sub0.

 For galaxies with SDSS-like observational noise shown in Fig.\ref{SDSS distance}, we see that the correlations of $f_{\rm innerhalo}$ versus $f_{\rm exsitu}$ are identical for galaxies at 40 Mpc with or without observational noise, and still the same when the galaxies are placed at 200 Mpc. However, it deviations for galaxies at larger distance, i.e. 400 Mpc or 600 Mpc. 
The inner stellar halo are contaminated by disk with PSF convolution. The luminosity fraction of the disk has a negative correlation with $f_{\rm exsitu}$, thus contamination by the disk could significantly diminish the ability to predict $f_{\rm exsitu}$ for the inner stellar halo. 

There is an offset in the correlation of $f_{\rm outerhalo}$ versus $f_{\rm exsitu}$ between those created from images with and without observation noise, while the correlations for galaxies with noise at 40-600 Mpc are similar. The luminosity of the outer halo is affected by the background noise; the good thing is that the effects of noise are stable for galaxies at different distances (here 40-600 Mpc). The other importance parameters, including $\nabla \rho_{\rm outer}$, $\nabla (\emph{g-r})_{\rm outer}$, $M_r$ and size $r_{\rm 90}$ are less affected by observational noise, and the correlations with $f_{\rm exsitu}$ are identical for galaxies at $d \lesssim 400$ Mpc, with or without observational noise.

We imposed a smaller PSF kernel and lower background noise for HSC-like mock images than for SDSS-like images. As shown in Fig. \ref {HSC distance}, with HSC-like observations, the correlation of $f_{\rm innerhalo}$ versus $f_{\rm exsitu}$ starts to deviate from the original for galaxies with $d \gtrsim 400$ Mpc. The correlation between $f_{\rm exsitu}$ and the other important parameters remains unchanged for galaxies with $d \lesssim 1000$ Mpc, and some of them, $\nabla \rho_{\rm outer}$ versus $f_{\rm exsitu}$ and $\nabla (\emph{g-r})_{\rm outer}$ versus $f_{\rm exsitu}$, begin to deviate from the original one for galaxies at $d=1500$ Mpc.

The inner halo luminosity $f_{\rm innerhalo}$ is easily affected by PSF, thus sensitive to the galaxy distance, our above model including $f_{\rm innerhalo}$ thus only capable for SDSS-like galaxies at $d\lesssim 200$ Mpc and HSC-like galaxies at $d\lesssim 400$ Mpc.

We created a new RF model with the combination of parameters Sub1 that are less affected by the galaxy distance: $\nabla \rho_{\rm outer}$ and $\nabla (\emph{g-r})_{\rm outer}$, $f_{\rm outerhalo}$,  $M_{\rm r}$, $r_{\rm 90}$, excluding $f_{\rm innerhalo}$ from Sub0. We train the model with $70\%$ of the galaxies at 40 Mpc, and make predictions for the other $30\%$ galaxies but put at further distances.
In Fig.\ref{fig:Sigmadistance}, we show the standard deviation of the model residuals ($\sigma(\Delta f_{\rm exsitu})$) for SDSS-like and HSC-like galaxies at difference distance. 

For galaxies at 40 Mpc, the residual $\sigma(\Delta f_{\rm exsitu})$ of the Sub1 model is similar to that of the model trained by Sub0 that includes $f_{\rm innerhalo}$. The surface brightness gradients $\nabla \rho_{\rm outer}$ contain information similar to $f_{\rm innerhalo}$ and $f_{\rm outerhalo}$, but are less affected by PSF. 
The residual of the model increases with the galaxy distance, but it still remains $\sigma(\Delta f_{\rm exsitu})\lesssim 0.1$ for SDSS-like galaxies at $r\lesssim 400$ Mpc ($z \lesssim 0.1$) and for HSC-like galaxies at $r\lesssim 1000$ Mpc ($z \lesssim 0.2$). Our model trained by the Sub1 parameters should be valid for galaxies in these regions. The future large surveys like LSST and CSST will observe a large sample of galaxies with high data quality, which will likely allow us to apply the model to galaxies at larger distances.

\subsection{Advantages and caveats}
\begin{table*}
\def\arraystretch{1.5}
\caption{Model performance compared with previous works.} 
\scriptsize\centering
\label{tab:compare}
\begin{tabular}{*{8}{l}}
\hline
       Model  & Data & Parameters & Stellar mass [$M_{\odot}$] & Redshift & Survey &$\sigma(\Delta f_{\rm exsitu})$  & Reference\\
       \hline
RF &TNG100 &  $\nabla \rho_{\rm outer}$, $\nabla (\emph{g-r})_{\rm outer}$, $f_{\rm outerhalo}$, $f_{\rm innerhalo}$, $M_{\rm r}$, $r_{\rm 90}$& $>10^{10.3}$ & z=0-0.2 & SDSS and HSC &0.08 & This work\\
RF &EAGLE &  - & - & - & - & 0.09 & This work\\
RF &TNG100 (train), &  - & - & - & - & 0.09 & This work\\&EAGLE (validate)\\
RF &EAGLE (train), &  - & - & - & - & 0.09 & This work\\&TNG100(validate)\\
RF &TNG100 &  $M_{*}$, $M_r$, $M_g$, $\emph{g-r}$, $r_{50}$, $r_{90}$, $C$, $\sigma$ & $>10^{10.16}$ & z=0 & SDSS & $\sim 0.1$ & \citep{2022MNRAS.515.3938S}\\
cINN  &TNG100 &  $M_{*}$, lookbacktime, $R_{e}$, $f_{\rm disk}$, $\emph{g-r}$, $Z_{*}$, $Age_{*}$ & $10^{10}-10^{12}$ & z=0-1 & No mock Survey& $\sim$0.06 & \citep{2023MNRAS.519.2199E}\\
CNN &TNG100 & 2D maps of mass, $v$, $\sigma$, age, metallicity within $1R_e$ & $>10^{10}$& z=0 & MaNGA& $\sim$0.07   & \citep{2023MNRAS.523.5408A}\\
CNN &EAGLE &  - & - & - & - & $\sim$0.08  & \citep{2023MNRAS.523.5408A}\\
CNN &TNG100 (train), &  - & - & - & - & $\sim$0.1  & \citep{2023MNRAS.523.5408A}\\&EAGLE(validate)\\
CNN &EAGLE (train), &  - & - & - & - & $\sim$0.1  & \citep{2023MNRAS.523.5408A}\\&TNG100(validate)\\

\hline
\end{tabular}
\tablefoot{In this work, we only use parameters directly derived from mock photometric images. In comparison, similar works in the literature include parameters or 2D maps of stellar mass, kinematics, age, and metallicity that can only be obtained from spectroscopic or even expensive IFU observations. }
\end{table*}

We summarise our model performance in Table~\ref{tab:compare}, for the four sets of models trained and validated by the TNG100 and EAGLE galaxies, or cross-validated, we have model residuals of $\sigma(\Delta f_{\rm exsitu})= 0.08-0.09$.
There are a few works in the literature trying to obtain galaxies' ex situ stellar mass fraction through the machine learning method. We compare with their results in Table~\ref{tab:compare}.

Most of the parameters used in \citet{2022MNRAS.515.3938S} are defined from mock photometric images like ours, including $M_{\rm r}$, $M_{\rm g}$, $r_{\rm 90}$, $r_{\rm 50}$, $\emph{g-r}$, concentration $C$. They included these parameters, as well as the total stellar mass $M_{*}$ and the single aperture velocity dispersion $\sigma$ to train a RF model. In their model, the most important parameters are the total stellar mass $M_{*}$ and the size $r_{\rm 90}$. Although they included $M_{*}$ as the input parameter, they still have $\sigma(\Delta f_{\rm exsitu}) \sim 0.1$, which is slightly larger than our model trained and validated by the same TNG100 galaxies. 
A better prediction with $\sigma(\Delta f_{\rm exsitu}) \sim 0.06$ is obtained by a cINN model from \cite{2023MNRAS.519.2199E}. However, they included some parameters that cannot be directly derived from photometric data and potentially could have a large uncertainty for real observations. The parameters they used include the total stellar mass $M_{*}$, the lookback time of the galaxy, half-light radius $R_{e}$, dynamically defined disk fraction $f_{\rm disk}$, luminosity-weighted stellar age $Age_{*}$ and metallicity $Z_{*}$. 

The above models were created by TNG100 galaxies and were not cross-validated by other simulations.
There is a CNN model using 2D maps of stellar mass, stellar velocity, velocity dispersion, stellar age, and metallicity that makes good predictions of $f_{\rm exsitu}$ with scatter of $\sim 0.07$ when trained and validated by the same TNG100 galaxies and was cross-validated by TNG100 and EAGLE galaxies \citep{2023MNRAS.523.5408A}. However, the input data used in this model can only be obtained from expensive IFU data.

Our model using only parameters defined from photometric data works similarly well to other models in the literature employing parameters that need spectroscopic observations or potentially harder to obtain. The luminosity fractions of the inner and outer halo ($f_{\rm innerhalo}$ and $f_{\rm outerhalo}$), or equivalently, the surface brightness and colour gradients ($\nabla \rho_{\rm outer}$ and $\nabla (g-r)_{\rm outer}$)) we defined play crucial importance here.

As any other models trained by simulations, our model could depend on the galaxy formation model and the resolution of the simulations. We cross-validated the model between TNG100, EAGLE, and also TNG50 (see Figures~\ref{fig3:appendix_tng50} and Figure~\ref{fig3:appendix_tng50fexSub}), which show that our model generally converges among these different simulations. However, the model prediction for some special type of galaxies could still be affected by the limited galaxy populations of the simulation. For example, EAGLE lacks galaxies with $f_{\rm exsitu}>0.7$, thus there is an upper limit of $f_{\rm exsitu}$ for TNG100 galaxies predicted by the EAGLE trained model as shown in the bottom left panel of Fig. \ref{fig:fex_fextrue}. 

A major limitation of our model is that we are only capable for edge-on galaxies. On the one hand, the definitions of edge-on galaxies from observations and simulations are different.  When applying the model to observations, it is hard to have a sample of galaxies perfectly edge-on, as theoretically defined. In the appendix, we show that our results are affected, but not so significantly, by the inclination angle. For a sample of edge-on galaxies mixed with 20\% of galaxies moderately inclined, our model still works statistically well. 
On the other hand, we lost a large fraction of real galaxies not edge-on.
For these galaxies not edge-on, the contamination of the disk will dilute the correlations and weaken the power of our model in predicting $f_{\rm exsitu}$. Potentially, the disk could be removed by photometric decomposition. However, photometric disk and bulge decomposition has not worked well for simulated galaxies \citep{2019MNRAS.483.4140R}, it may not lead to consistent results with real observations. We may further validate the method by trying photometric decomposition to observed non-edge-on galaxies, and compare with the results with edge-on galaxies.

%---------------------------------------------------
\section{Conclusion}
%---------------------------------------------------
 We created mock images using galaxies in the cosmological hydrodynamical simulations TNG100, EAGLE, and TNG50 at redshift $z=0$. We projected all galaxies as edge-on, and defined a series of parameters describing their structures, including: the absolute magnitude in $r$ and $g$ bands ($M_r$, $M_g$), the half-light and 90\%-light radius ($r_{50}$, $r_{90}$), the concentration ($C$), the luminosity fractions of inner and outer halos ($f_{\rm innerhalo}$, $f_{\rm outerhalo}$), the inner and outer surface brightness gradient ($\nabla \rho_{\rm inner}$ and $\nabla \rho_{\rm outer}$), color gradients ($\nabla (g-r)_{\rm inner}$ and $\nabla (g-r)_{\rm outer}$) and the single-aperture velocity dispersion $\sigma$ and Kurtosis from Gaussian-Hermit fitting ($h_4$). 
 
 In particular, the inner and outer halo of a galaxy are defined by a sector along the minor axis, with $45-135$ degrees from the disk major axis, and with radii ranging from $3.5-10$ kpc and $10-30$ kpc, respectively, to avoid contamination of disk and bulge. The surface brightness and the colour gradients are defined in the same sector along the minor axis, and in radii ranges of 1.5-10 kpc and 10-30 kpc for the inner and outer gradients, respectively.
 We then evaluated the importance of these parameters in predicting the ex situ stellar mass fraction $f_{\rm exsitu}$, and constructed machine learning models using the random forest(RF) method to predict $f_{\rm exsitu}$. 

Our main results are as follows.
\begin{enumerate}
\item We find that the outer gradients of surface brightness ($\nabla \rho_{\rm outer}$) and colour ($\nabla (g-r)_{\rm outer}$), as well as the luminosity fraction of the inner and outer halo ($f_{\rm innerhalo}$ and $f_{\rm outerhalo}$) are strongly correlated with the ex situ stellar mass fraction $f_{\rm exsitu}$, and these correlations are almost identical in TNG100, EAGLE, and TNG50, independent of the galaxy formation model and simulation resolution. 

\item We evaluate the importance of all structure parameters in predicting the ex situ stellar mass fraction ($f_{\rm exsitu}$). We find that the parameters of highest importance in the model are $\nabla \rho_{\rm outer}$, $\nabla \rm (\emph{g-r})_{outer}$, $f_{\rm outerhalo}$, and $f_{\rm innerhalo}$, and followed by the absolute magnitude $M_r$, the galaxy size $r_{\rm 90}$; the remaining parameters are not important.

\item We train RF models by including the six most important parameters (Sub0: $\nabla \rho_{\rm outer}$, $\nabla \rm (\emph{g-r})_{outer}$, $f_{\rm outerhalo}$, $f_{\rm innerhalo}$, $M_{\rm r}$, $r_{\rm 90}$) measured from mock photometric images: the models effectively predict the ex situ fraction with residual $\sigma(\Delta f_{\rm exsitu})< 0.1$. The models trained from TNG100 and EAGLE work similarly well and are cross-validated; they also work well in making predictions for TNG50 galaxies.

\item The correlations of $\nabla \rho_{\rm outer}$ versus $f_{\rm exsitu}$, $\nabla \rm (\emph{g-r})_{outer}$ versus $f_{\rm exsitu}$ and $f_{\rm outerhalo}$ versus $f_{\rm exsitu}$ are affected by observational noise, but remain unchanged for galaxies with $d\lesssim 400$ Mpc ($z \lesssim 0.1$) for SDSS-like observations and $d\lesssim 1000$ Mpc ($z \lesssim 0.2$) for HSC-like observations. Our model is thus validated for galaxies with such data quality and within these distances.
\end{enumerate}

In summary, the luminosity and colour gradients, as well as the luminosity fraction of the inner and outer halo, can robustly predict the ex situ stellar mass fraction in nearby galaxies. The correlations between these parameters and $f_{\rm exsitu}$ are almost identical in all the simulations we explore. Our models trained by RF can be transferred between TNG100, EAGLE, and TNG50 galaxies, thus could potentially be applied to real galaxies from large photometric surveys.

\label{6}

%----------------------------------------------------

% ----------------------------------------------------

\begin{acknowledgement}
The authors thank Song Huang for useful discussions.
LZ acknowledges the support of the National Key R\&D Programme of China No. 2022YFF0503403, and the CAS Project for Young Scientists in Basic Research under grant No. YSBR-062. WW acknowledge the support of NSFC (12273021), the National Key R\&D Programme of China (2023YFA1605600, 2023YFA1605601) and the Yangyang Development Fund. JF-B acknowledges support from the PID2022-140869NB-I00 grant from the Spanish Ministry of Science and Innovation.
\end{acknowledgement}

%---------------------------------------------------------------------
\bibliographystyle{aa}  % style aa.bst
\bibliography{ms_relic} % your references Yourfile.bib
%---------------------------------------------------------------------

%---------------------------------------------------------------------
%\usepackage{appendix}
\begin{appendix}\onecolumn
\section{Galaxies with ongoing mergers}
\begin{figure*}[h!]
\centering
\subfloat{\includegraphics[width=6cm]{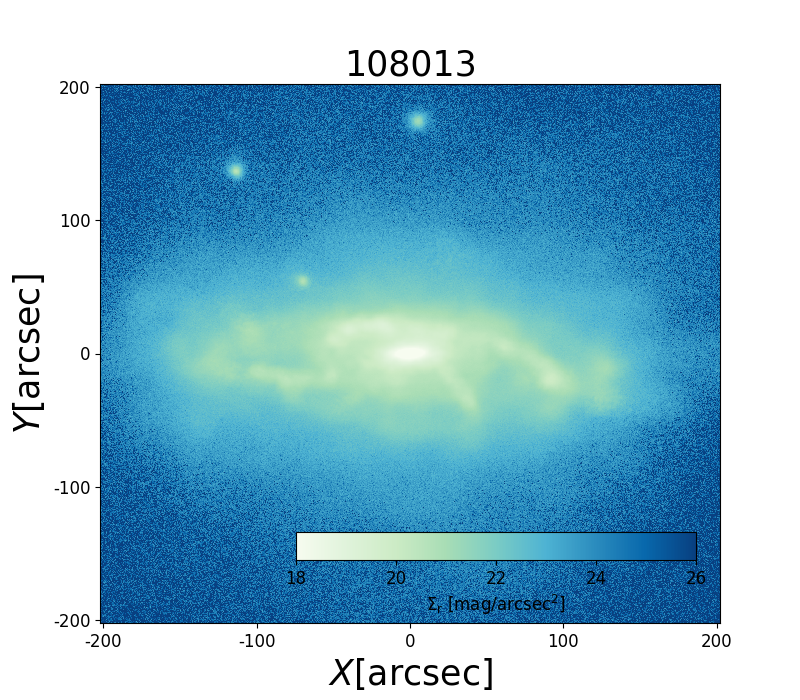}\includegraphics[width=6cm]{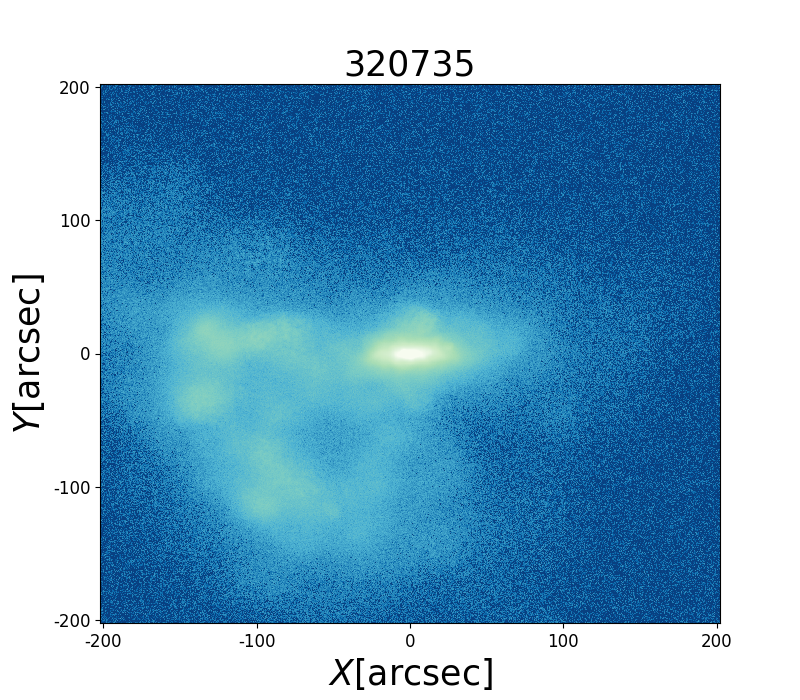}\includegraphics[width=6cm]{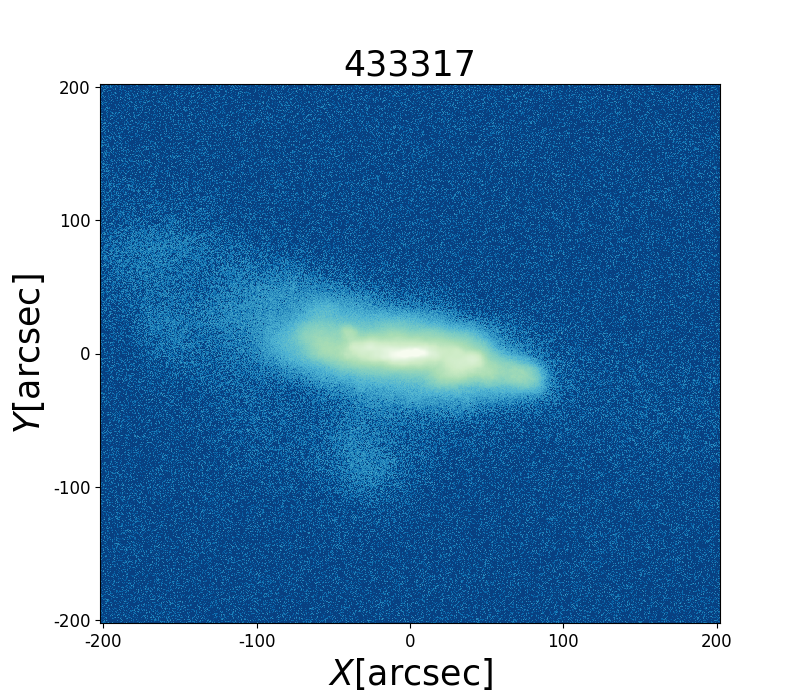}}
\quad
\subfloat{\includegraphics[width=6cm]{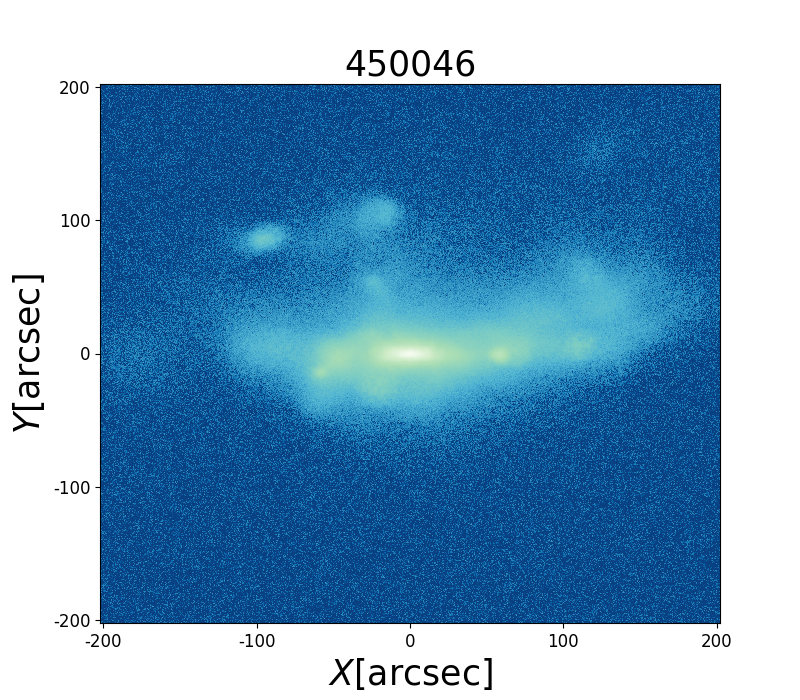}\includegraphics[width=6cm]{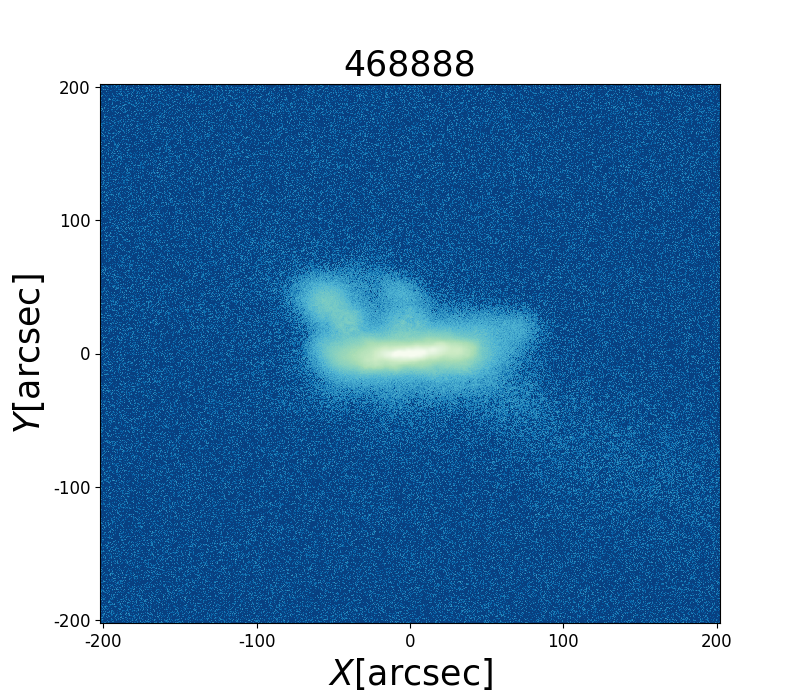}\includegraphics[width=6cm]{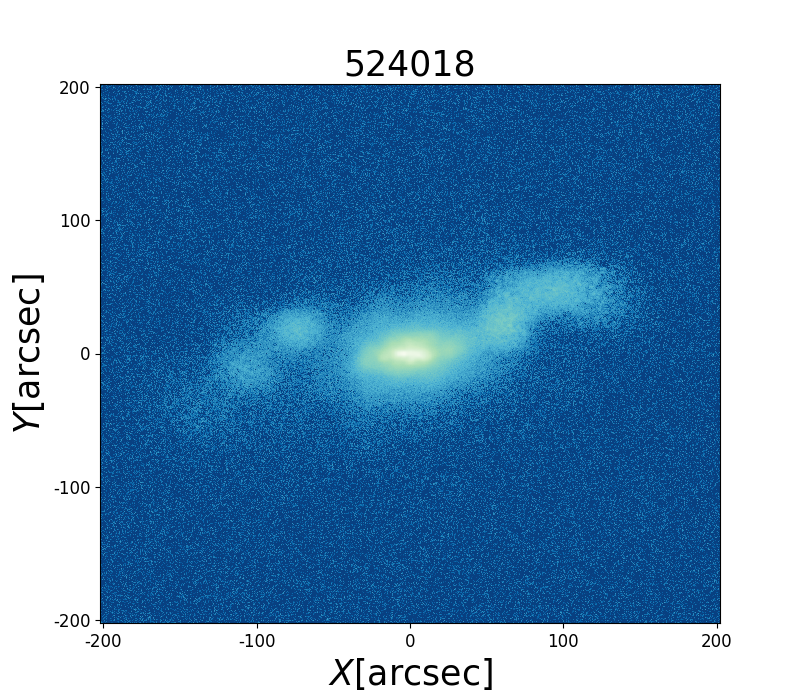}}
\caption{Six cases of galaxies with ongoing mergers, with their subhaloID labeled at the top.
}
\label{fig:ongoing_merger}
\end{figure*}

We check all the SDSS-like mock images by eyes, and identify these galaxies with obvious substructures in the inner halo or with disk obviously disturbed as ongoing merging galaxies. These galaxies are excluded in our sample. We show six of such cases in Fig.\ref{fig:ongoing_merger}. 

\section{Model prediction for TNG50 galaxies}

\begin{figure*}[h!]
\centering
\flushleft\subfloat{
		\includegraphics[scale=0.45]{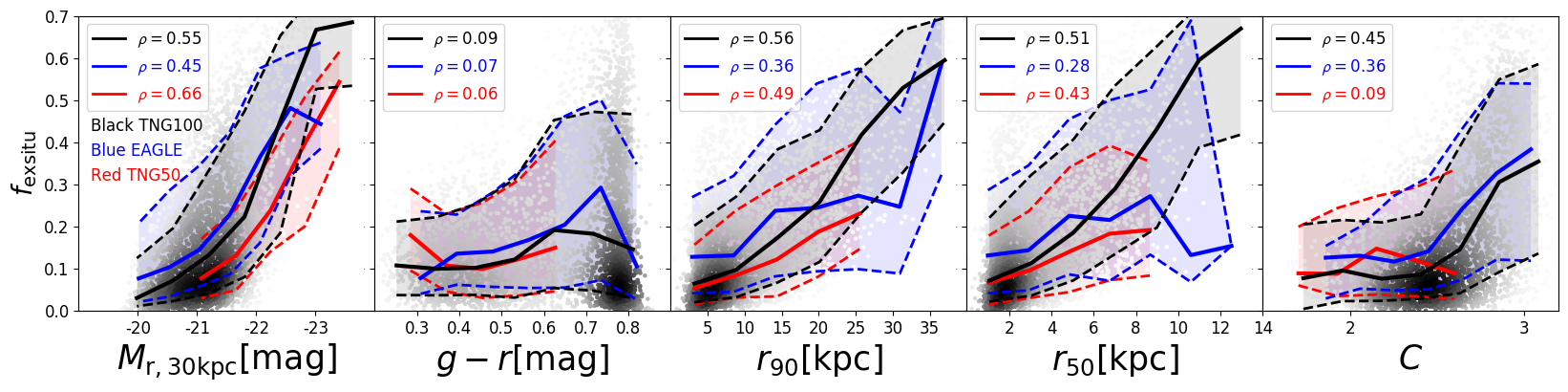}}
\flushleft\subfloat{
		\includegraphics[scale=0.45]{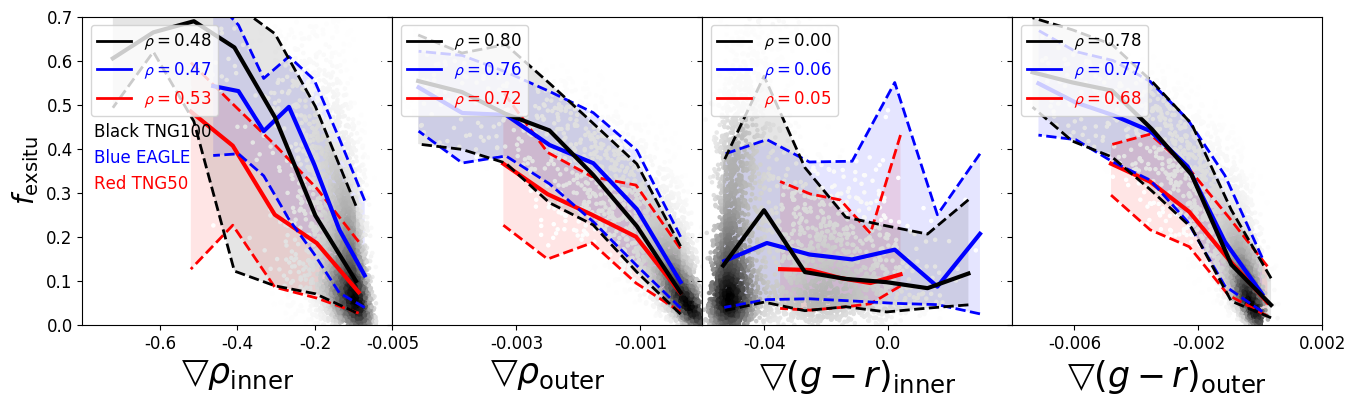}}
\flushleft\subfloat{
		\includegraphics[scale=0.45]{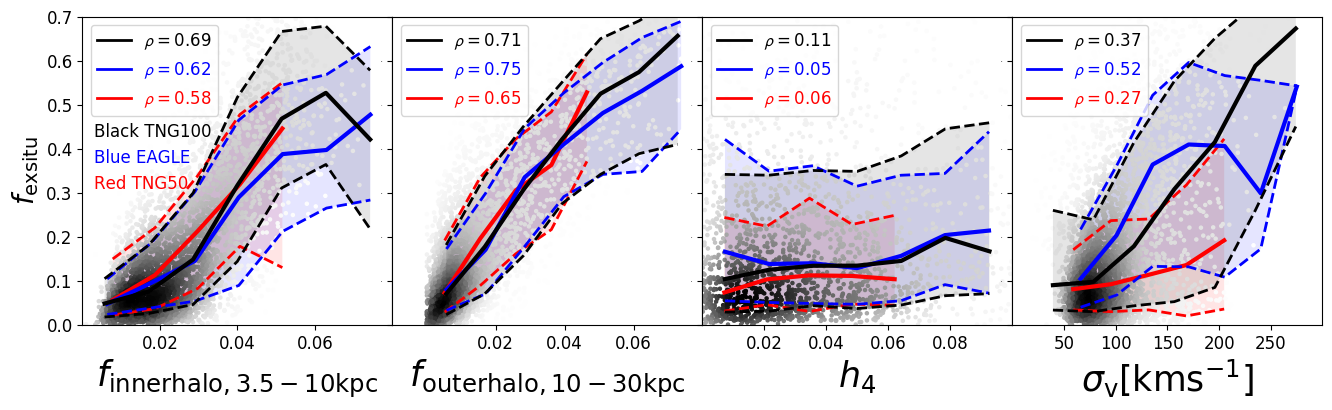}}
\caption{The correlation of ex situ stellar mass fraction $f_{\rm exsitu}$ with morphological parameters. The figure is similar to Figure \ref{fig3:fexsitufeature} but with the morphological parameters derived from the clean images without observational noise. The black, blue, and red solid curves are running median for TNG100, EAGLE, and TNG50 galaxies, the dashed curves represent the ±1$\sigma$ scatter.
}
\label{fig3:appendix_tng50}
\end{figure*}

In Figure \ref{fig3:appendix_tng50}, we show that the TNG50 galaxies have trends similar to those of TNG100 and EAGLE in the correlations of $f_{\rm innerhalo}$, $f_{\rm outerhalo}$, $\nabla \rho_{\rm outer}$ and $\nabla \rm (\emph{g-r})_{outer}$ versus $f_{\rm exsitu}$, and they have obvious offsets with TNG100 and EAGLE in the other correlations. The parameters $f_{\rm innerhalo}$, $f_{\rm outerhalo}$, $\nabla \rho_{\rm outer}$ and $\nabla \rm (\emph{g-r})_{outer}$ are probably not affected by simulation resolutions. 

\begin{figure*}[h!]
\centering
\subfloat{\includegraphics[width=7cm]{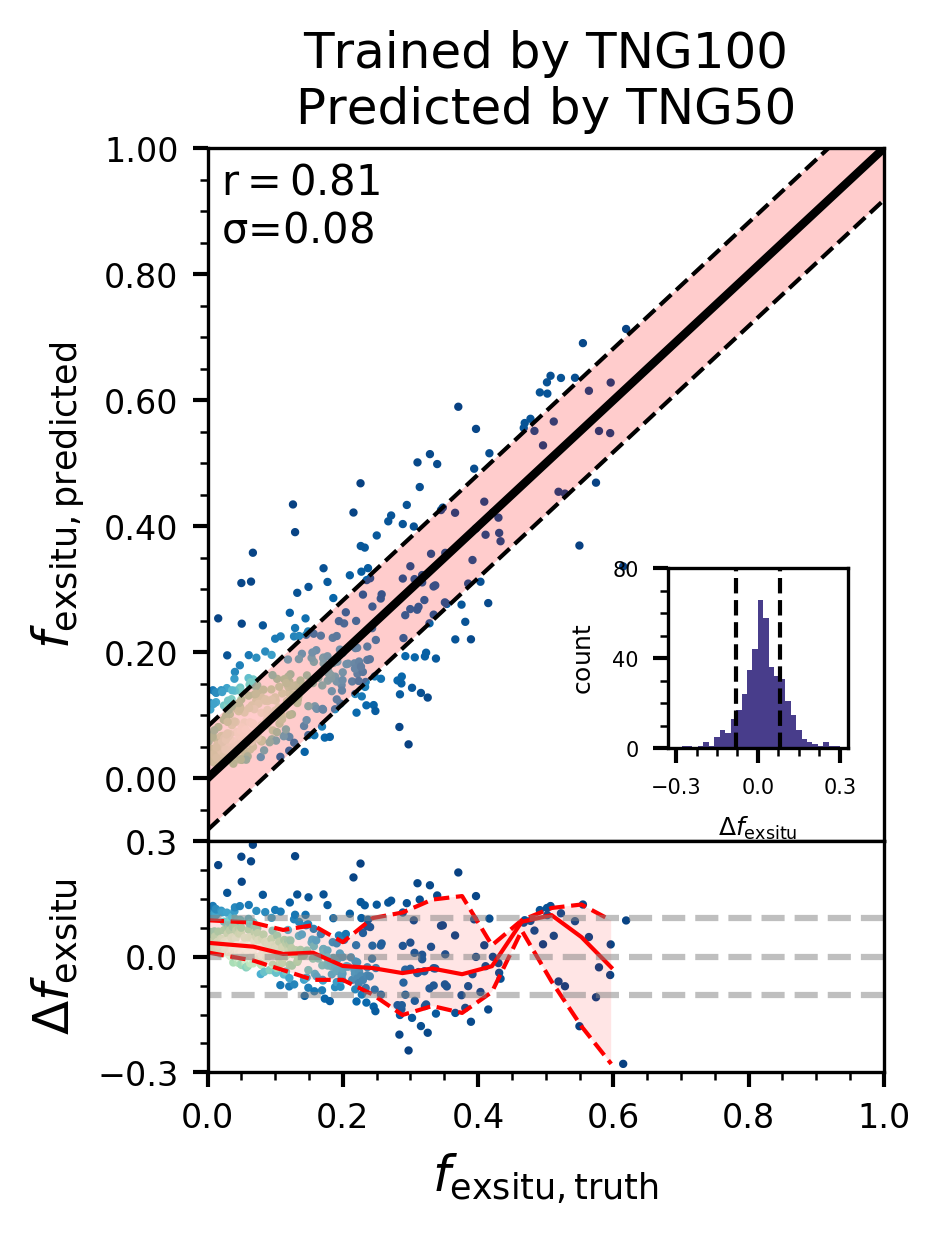}\includegraphics[width=7cm]{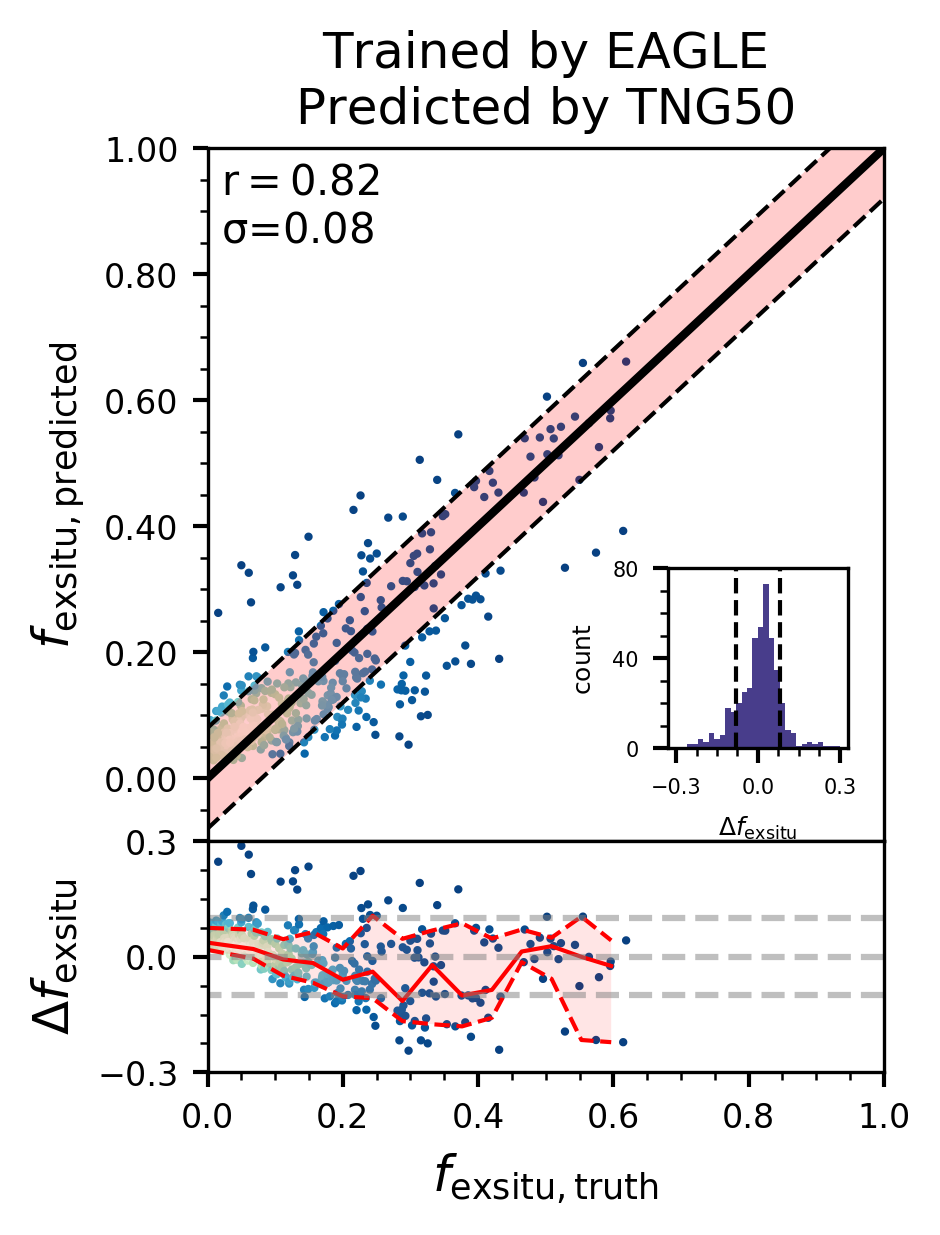}}
\caption{Model predictions versus ground truth, similar with Fig.\ref{fig:fex_fextrue}, but making predictions for TNG50 galaxies with clean images.}
\label{fig3:appendix_tng50fexSub}
\end{figure*}

To further check if the simulation resolution will affect our results, we use the models trained by the mock data created from TNG100 and EAGLE, and make predictions for TNG50 galaxies. We show the model prediction versus the ground truth of $f_{\rm exsitu}$ in Figure \ref{fig3:appendix_tng50fexSub}, the models work well with $\sigma(\Delta(f_{\rm exsitu})) = 0.08$. Our model probably already converges at the resolution of TNG100.

\section{The effects of inclination angle on the model prediction}

We limited our model to edge-on galaxies in the paper. However, the definitions of edge-on galaxies from observations and simulations are different. When applying the model to observations, it is hard to have a sample of galaxies perfectly edge-on, as theoretically defined. Here we evaluate how the results will be affected if the sample is mixed with some galaxies not perfectly edge-on.

\begin{figure*}[h!]
\centering
\subfloat{\includegraphics[width=7cm]{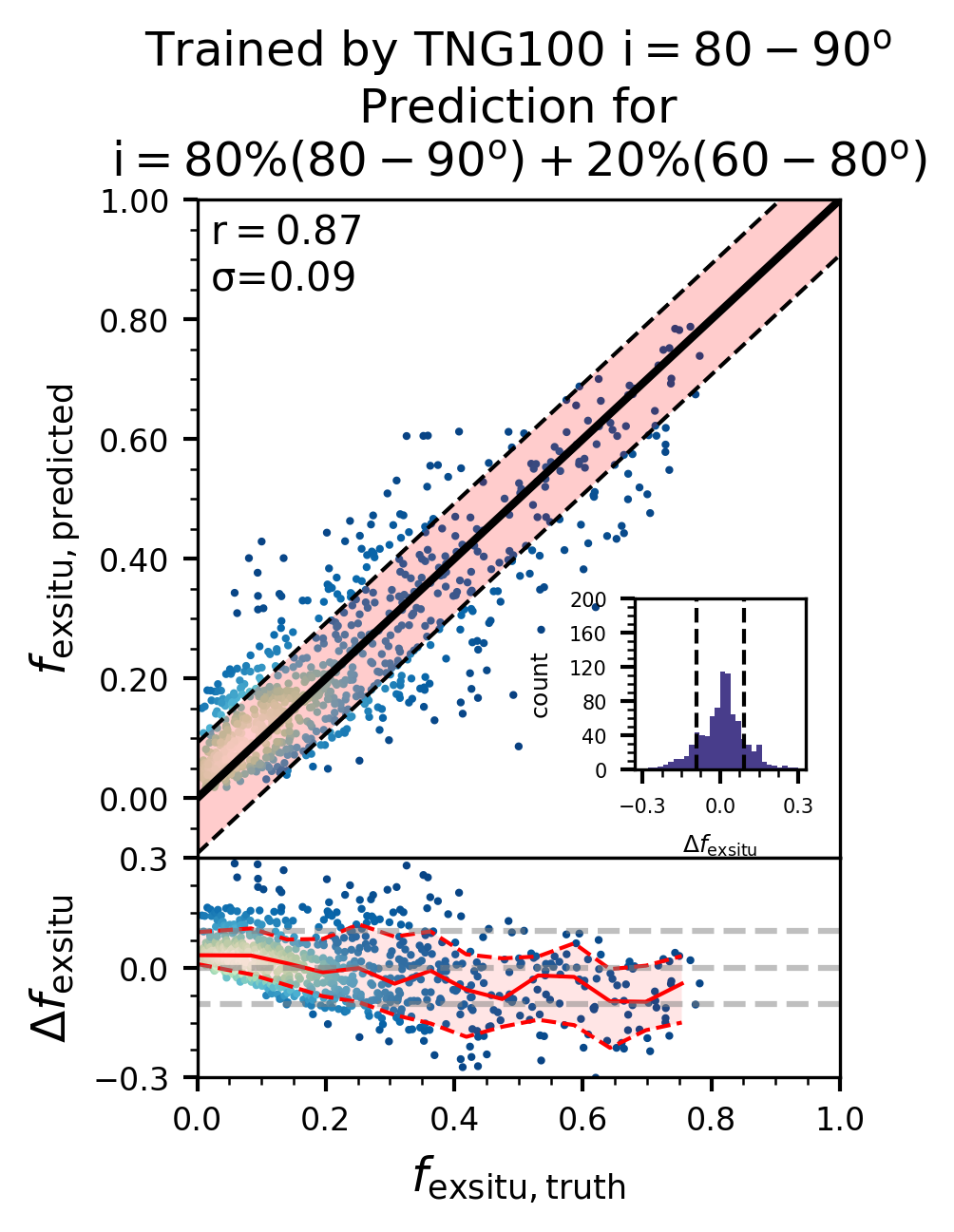}\includegraphics[width=7cm]{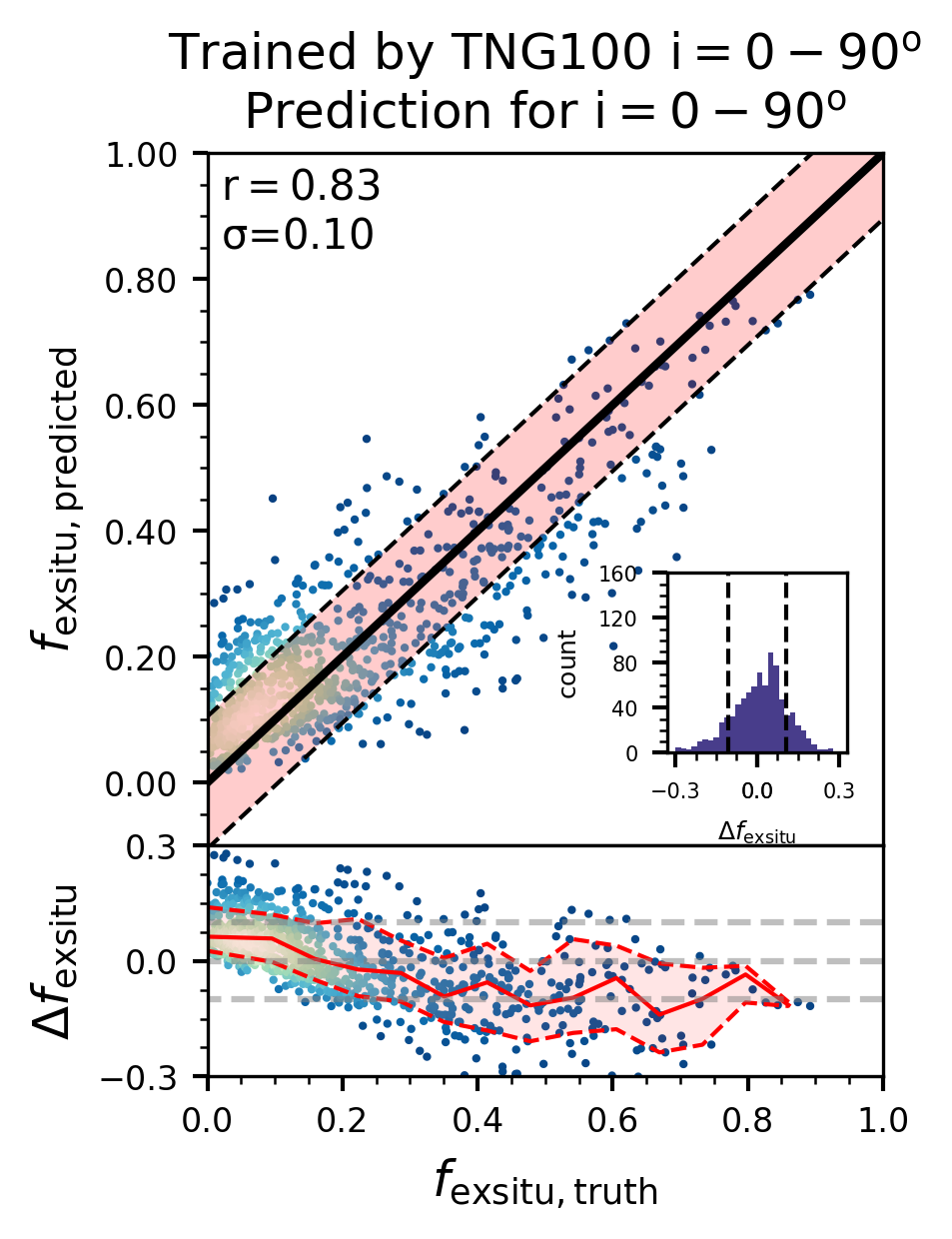}}
\caption{Model predictions versus ground truth, similar with Fig. \ref{fig:fex_fextrue}. Left panel: model trained by TNG100 galaxies with $i = 80^o-90^o$ but making prediction for a sample composed of 80\% galaxies with $i = 80^o-90^o$ and mixed with 20\% with $i = 60^o-80^o$. Right panel: Model trained and validated by TNG100 galaxies randomly projected between $i = 0^o-90^o$.}
\label{fig:mix20precent60-80}
\end{figure*}

 We used the model trained by TNG100 edge-on galaxies shown in the paper (inclination angle $i = 80^o-90^o$), and made predictions for a sample composed of 80\% galaxies with $i = 80^o-90^o$ and mixed with 20\% with $i = 60^o-80^o$. In Figure~\ref{fig:mix20precent60-80}, we show the the model prediction versus ground truth. The model works statistically well for the mixed sample, with $\mathcal{R} = 0.87$ and the standard deviation of $\sigma(\Delta(f_{\rm exsitu})) = 0.09$, similar to the model prediction for all edge-on galaxies. 
 
 We also trained and validated a model using TNG100 galaxies randomly projected between $i = 0^o-90^o$. As shown in the right panel of Figure~\ref{fig:mix20precent60-80}, the model works less well compared to that limited to edge-on galaxies. Especially for the disk galaxies with low ex situ stellar mass, the model tends to over-predict their ex situ stellar mass fractions.

\end{appendix}
\end{document}